\def\submit{1}

\documentclass[12pt]{article}

\newcommand{\Papertitle}{Low Rank Alternating Direction Method of Multipliers Reconstruction for MR Fingerprinting}

\usepackage{forloop}
\usepackage{xcolor}
\definecolor{UKLred} {RGB}{207, 25,  59}
\definecolor{UKLblue}{RGB}{ 47, 63, 157}
\definecolor{turquois}{rgb}{0,0.75,0.75}%
	
\usepackage{soul} 								
\usepackage[textsize=tiny]{todonotes}				

\usepackage[english]{babel}
\usepackage[sc]{mathpazo}
\usepackage[T1]{fontenc}
\usepackage{textcomp}
\usepackage[utf8]{inputenc}

\usepackage{graphicx} 

\usepackage{amsmath}
\usepackage{amssymb}
\usepackage{amsthm}
\usepackage{amsfonts}
\usepackage{upgreek}

\usepackage{helvet}

\usepackage[helvet]{sfmath}

\usepackage{multirow}
\usepackage{dsfont}

\if\submit0
\usepackage{tikz}
\usepackage{pgfplots}
\pgfplotsset{compat=1.9,
		colormap={morgenstemning}{[1pt] rgb(0pt)=(0,0,0); rgb(1pt)=(0.000300784,0.00401529,0.00572235); rgb(2pt)=(0.000501569,0.00793059,0.0114447); rgb(3pt)=(0.000703529,0.0118459,0.0171682); rgb(4pt)=(0.00100471,0.0157596,0.0229894); rgb(5pt)=(0.00130588,0.0195765,0.0287137); rgb(6pt)=(0.00160941,0.0234918,0.0345341); rgb(7pt)=(0.00201098,0.0274043,0.0402592); rgb(9pt)=(0.00281412,0.0350341,0.0519047); rgb(10pt)=(0.00321961,0.038849,0.0577275); rgb(11pt)=(0.00372157,0.0426639,0.0635502); rgb(12pt)=(0.00422353,0.0464741,0.0693729); rgb(13pt)=(0.00473059,0.0501886,0.0751957); rgb(14pt)=(0.00533294,0.0539031,0.0810239); rgb(15pt)=(0.00593529,0.0576176,0.0869471); rgb(16pt)=(0.00654392,0.0613259,0.0928702); rgb(17pt)=(0.00724667,0.06494,0.0987933); rgb(18pt)=(0.00794941,0.0685471,0.104716); rgb(19pt)=(0.00865961,0.0720682,0.11064); rgb(20pt)=(0.00946275,0.0756824,0.116563); rgb(21pt)=(0.0102741,0.0792882,0.122486); rgb(22pt)=(0.0111776,0.082802,0.128409); rgb(23pt)=(0.0120812,0.0863157,0.134341); rgb(24pt)=(0.0129941,0.0898294,0.140355); rgb(25pt)=(0.013998,0.0933333,0.146278); rgb(26pt)=(0.015002,0.0967569,0.152212); rgb(27pt)=(0.0160165,0.10026,0.158225); rgb(28pt)=(0.0171208,0.103684,0.164159); rgb(29pt)=(0.0182365,0.107187,0.170171); rgb(31pt)=(0.0206459,0.114013,0.182017); rgb(32pt)=(0.0218506,0.117439,0.18794); rgb(33pt)=(0.0230682,0.120953,0.193851); rgb(34pt)=(0.0243733,0.124467,0.199687); rgb(35pt)=(0.0256784,0.12798,0.205596); rgb(36pt)=(0.0269976,0.131494,0.211433); rgb(37pt)=(0.0284031,0.134993,0.217342); rgb(38pt)=(0.0298086,0.138407,0.223179); rgb(39pt)=(0.0312294,0.14182,0.229087); rgb(40pt)=(0.032751,0.145218,0.234925); rgb(41pt)=(0.0343733,0.148531,0.240849); rgb(42pt)=(0.03608,0.151844,0.246772); rgb(43pt)=(0.0378035,0.15514,0.252712); rgb(44pt)=(0.0396278,0.158335,0.258718); rgb(45pt)=(0.0415353,0.161465,0.264641); rgb(46pt)=(0.0434608,0.164659,0.270564); rgb(47pt)=(0.0454871,0.16779,0.276487); rgb(48pt)=(0.0475953,0.171002,0.282392); rgb(49pt)=(0.0497227,0.174215,0.288215); rgb(50pt)=(0.051951,0.177408,0.294037); rgb(51pt)=(0.05428,0.18052,0.29986); rgb(52pt)=(0.0567098,0.183591,0.305683); rgb(53pt)=(0.0592612,0.186503,0.311526); rgb(54pt)=(0.0619929,0.189372,0.317449); rgb(55pt)=(0.0648255,0.192082,0.323373); rgb(56pt)=(0.0677808,0.194793,0.329296); rgb(57pt)=(0.0709153,0.197459,0.335196); rgb(58pt)=(0.0741961,0.199946,0.341065); rgb(59pt)=(0.0777561,0.202286,0.347065); rgb(60pt)=(0.0815412,0.204371,0.353012); rgb(61pt)=(0.0856286,0.206307,0.359035); rgb(62pt)=(0.0900427,0.207916,0.365059); rgb(63pt)=(0.0949094,0.209098,0.371132); rgb(64pt)=(0.100456,0.209725,0.377356); rgb(65pt)=(0.106884,0.2098,0.383529); rgb(66pt)=(0.114041,0.209774,0.389449); rgb(67pt)=(0.121728,0.209674,0.395019); rgb(68pt)=(0.129967,0.209547,0.40036); rgb(69pt)=(0.138608,0.209319,0.405399); rgb(70pt)=(0.14768,0.20899,0.410135); rgb(71pt)=(0.157201,0.208589,0.414597); rgb(72pt)=(0.167024,0.208159,0.418829); rgb(73pt)=(0.17712,0.207657,0.422788); rgb(74pt)=(0.187447,0.207126,0.426487); rgb(75pt)=(0.198076,0.206494,0.429841); rgb(76pt)=(0.209008,0.205762,0.432964); rgb(77pt)=(0.220212,0.204928,0.435815); rgb(78pt)=(0.231587,0.203994,0.438434); rgb(79pt)=(0.243094,0.202959,0.440782); rgb(80pt)=(0.254739,0.201824,0.442927); rgb(81pt)=(0.266385,0.200619,0.444904); rgb(82pt)=(0.278095,0.19935,0.446715); rgb(83pt)=(0.289941,0.197944,0.448288); rgb(84pt)=(0.30182,0.196506,0.449695); rgb(85pt)=(0.313767,0.194933,0.450867); rgb(86pt)=(0.325747,0.193193,0.451904); rgb(87pt)=(0.337794,0.191352,0.452773); rgb(88pt)=(0.349876,0.18941,0.453507); rgb(89pt)=(0.361988,0.187332,0.454075); rgb(90pt)=(0.374035,0.185124,0.454541); rgb(91pt)=(0.386047,0.182844,0.454907); rgb(92pt)=(0.397957,0.180398,0.455172); rgb(93pt)=(0.409804,0.177888,0.455373); rgb(94pt)=(0.421613,0.175378,0.455537); rgb(95pt)=(0.433284,0.172831,0.4556); rgb(96pt)=(0.444792,0.170259,0.4556); rgb(97pt)=(0.456198,0.167787,0.455562); rgb(98pt)=(0.467504,0.165378,0.455462); rgb(99pt)=(0.478709,0.162968,0.455322); rgb(100pt)=(0.489814,0.160559,0.455122); rgb(101pt)=(0.500778,0.15811,0.454921); rgb(102pt)=(0.51162,0.15564,0.45472); rgb(103pt)=(0.522422,0.153231,0.454479); rgb(104pt)=(0.533123,0.150821,0.454178); rgb(105pt)=(0.543724,0.148412,0.453835); rgb(106pt)=(0.554265,0.146002,0.453434); rgb(107pt)=(0.564764,0.143593,0.453032); rgb(108pt)=(0.575162,0.141226,0.452588); rgb(109pt)=(0.58546,0.138917,0.452086); rgb(110pt)=(0.5957,0.136608,0.451584); rgb(111pt)=(0.605853,0.134299,0.451082); rgb(112pt)=(0.615892,0.13199,0.450624); rgb(113pt)=(0.625887,0.129681,0.450178); rgb(114pt)=(0.635781,0.127372,0.449721); rgb(115pt)=(0.645575,0.125108,0.449275); rgb(116pt)=(0.655313,0.122945,0.448773); rgb(117pt)=(0.665005,0.120882,0.448271); rgb(118pt)=(0.674596,0.118967,0.447769); rgb(119pt)=(0.684087,0.11716,0.44722); rgb(120pt)=(0.693524,0.1154,0.446665); rgb(121pt)=(0.702913,0.113741,0.446115); rgb(122pt)=(0.712249,0.11223,0.445513); rgb(123pt)=(0.721538,0.110921,0.444911); rgb(124pt)=(0.730725,0.109765,0.44426); rgb(125pt)=(0.73991,0.108759,0.443508); rgb(126pt)=(0.749096,0.107954,0.442655); rgb(127pt)=(0.758282,0.107451,0.441652); rgb(128pt)=(0.767518,0.108304,0.440096); rgb(129pt)=(0.776704,0.111273,0.437785); rgb(130pt)=(0.78589,0.115851,0.43492); rgb(131pt)=(0.795024,0.121739,0.431499); rgb(132pt)=(0.804007,0.128627,0.427681); rgb(133pt)=(0.812942,0.136377,0.423557); rgb(134pt)=(0.821719,0.145187,0.418978); rgb(135pt)=(0.8303,0.154894,0.414053); rgb(136pt)=(0.838673,0.165093,0.408927); rgb(137pt)=(0.846852,0.175702,0.403599); rgb(138pt)=(0.854821,0.186924,0.398015); rgb(139pt)=(0.862543,0.19885,0.392129); rgb(140pt)=(0.869953,0.211537,0.385886); rgb(141pt)=(0.877071,0.224865,0.379351); rgb(142pt)=(0.883931,0.23864,0.372558); rgb(143pt)=(0.890489,0.252707,0.365518); rgb(144pt)=(0.896801,0.266806,0.358391); rgb(145pt)=(0.902912,0.280918,0.351206); rgb(146pt)=(0.908764,0.295187,0.343863); rgb(147pt)=(0.914313,0.309659,0.336376); rgb(148pt)=(0.91956,0.32439,0.328673); rgb(149pt)=(0.924505,0.339423,0.320784); rgb(150pt)=(0.929147,0.3547,0.312735); rgb(151pt)=(0.933428,0.370178,0.304485); rgb(152pt)=(0.937365,0.385918,0.295974); rgb(153pt)=(0.941,0.4019,0.28728); rgb(154pt)=(0.944333,0.417963,0.278425); rgb(155pt)=(0.947424,0.434086,0.269429); rgb(156pt)=(0.950252,0.450249,0.260272); rgb(157pt)=(0.952778,0.466351,0.251036); rgb(158pt)=(0.955063,0.482352,0.241738); rgb(159pt)=(0.957209,0.498189,0.232401); rgb(160pt)=(0.959192,0.513763,0.223002); rgb(161pt)=(0.961036,0.529097,0.213565); rgb(162pt)=(0.962716,0.544229,0.204128); rgb(163pt)=(0.964259,0.559224,0.194755); rgb(164pt)=(0.965636,0.57399,0.185419); rgb(165pt)=(0.966876,0.588582,0.176082); rgb(166pt)=(0.967951,0.603009,0.166811); rgb(167pt)=(0.968889,0.617234,0.15764); rgb(168pt)=(0.969727,0.631258,0.148636); rgb(169pt)=(0.970464,0.64508,0.139768); rgb(170pt)=(0.971033,0.6587,0.131); rgb(171pt)=(0.971535,0.672052,0.1224); rgb(172pt)=(0.972037,0.685169,0.113934); rgb(173pt)=(0.972471,0.698084,0.105637); rgb(174pt)=(0.972873,0.710729,0.0974729); rgb(175pt)=(0.973275,0.723141,0.0894098); rgb(176pt)=(0.973676,0.735351,0.0813094); rgb(177pt)=(0.974147,0.747291,0.0732471); rgb(178pt)=(0.974579,0.758927,0.0653553); rgb(179pt)=(0.974981,0.770402,0.0578055); rgb(180pt)=(0.975382,0.781635,0.0506588); rgb(181pt)=(0.975784,0.792637,0.0439153); rgb(182pt)=(0.976257,0.803437,0.0375749); rgb(183pt)=(0.976759,0.814035,0.0315659); rgb(184pt)=(0.977333,0.82436,0.0257149); rgb(185pt)=(0.978008,0.834455,0.0202549); rgb(186pt)=(0.978711,0.844348,0.0153718); rgb(187pt)=(0.979487,0.854113,0.0110933); rgb(188pt)=(0.980364,0.86363,0.00727216); rgb(189pt)=(0.981267,0.872993,0.00407647); rgb(190pt)=(0.982245,0.882255,0.00166078); rgb(191pt)=(0.983249,0.891416,0.000276078); rgb(192pt)=(0.984178,0.900627,0.00240941); rgb(193pt)=(0.985081,0.909787,0.00811961); rgb(194pt)=(0.985985,0.918695,0.0162427); rgb(195pt)=(0.986965,0.9273,0.0265588); rgb(196pt)=(0.987969,0.935449,0.0382467); rgb(197pt)=(0.988973,0.943249,0.0514294); rgb(198pt)=(0.989976,0.950668,0.0663024); rgb(199pt)=(0.990902,0.957607,0.0824204); rgb(200pt)=(0.991727,0.964276,0.0987843); rgb(201pt)=(0.992531,0.970508,0.1157); rgb(202pt)=(0.993413,0.976157,0.133559); rgb(203pt)=(0.994237,0.981201,0.152425); rgb(204pt)=(0.99496,0.98572,0.17238); rgb(205pt)=(0.995663,0.989494,0.193282); rgb(206pt)=(0.996365,0.992543,0.215092); rgb(207pt)=(0.997149,0.994805,0.237808); rgb(208pt)=(0.997953,0.996179,0.261595); rgb(209pt)=(0.998592,0.996892,0.28608); rgb(210pt)=(0.998865,0.997247,0.310841); rgb(211pt)=(0.9989,0.9973,0.335638); rgb(212pt)=(0.998651,0.997051,0.360352); rgb(213pt)=(0.998182,0.996666,0.384547); rgb(214pt)=(0.997596,0.996264,0.408054); rgb(215pt)=(0.997163,0.995863,0.430349); rgb(216pt)=(0.997015,0.9958,0.450572); rgb(217pt)=(0.997085,0.99597,0.469428); rgb(218pt)=(0.997185,0.996171,0.487503); rgb(219pt)=(0.997286,0.996458,0.504958); rgb(220pt)=(0.997473,0.996759,0.522067); rgb(221pt)=(0.997587,0.996973,0.538787); rgb(222pt)=(0.997774,0.997261,0.555104); rgb(223pt)=(0.997887,0.997475,0.57128); rgb(224pt)=(0.9979,0.997588,0.587443); rgb(225pt)=(0.9979,0.9976,0.603429); rgb(226pt)=(0.9979,0.997689,0.619215); rgb(227pt)=(0.9979,0.9977,0.634709); rgb(228pt)=(0.9979,0.997789,0.649901); rgb(229pt)=(0.9979,0.9978,0.664881); rgb(230pt)=(0.99799,0.99789,0.679659); rgb(231pt)=(0.998,0.9979,0.694145); rgb(232pt)=(0.998091,0.997991,0.708328); rgb(233pt)=(0.998191,0.998183,0.722301); rgb(234pt)=(0.998292,0.998292,0.736072); rgb(235pt)=(0.998484,0.998484,0.749641); rgb(236pt)=(0.998685,0.998685,0.763009); rgb(237pt)=(0.998793,0.998793,0.776268); rgb(238pt)=(0.998987,0.998893,0.789427); rgb(239pt)=(0.999094,0.998994,0.802578); rgb(240pt)=(0.999194,0.999094,0.815729); rgb(241pt)=(0.9992,0.9991,0.828786); rgb(242pt)=(0.9992,0.9991,0.841742); rgb(243pt)=(0.9992,0.9991,0.854693); rgb(244pt)=(0.9992,0.9991,0.867452); rgb(245pt)=(0.9992,0.9991,0.880106); rgb(246pt)=(0.9992,0.9991,0.892659); rgb(247pt)=(0.9992,0.999197,0.905014); rgb(248pt)=(0.999297,0.999297,0.917265); rgb(249pt)=(0.9993,0.9993,0.929415); rgb(250pt)=(0.999398,0.999398,0.941465); rgb(251pt)=(0.999498,0.999498,0.953413); rgb(252pt)=(0.999599,0.999599,0.965162); rgb(253pt)=(0.999699,0.999699,0.976908); rgb(254pt)=(0.9998,0.9998,0.988555); rgb(255pt)=(1,1,1)},
		colormap={hotblack}{[1pt] color(0pt)=(blue); color(499pt)=(yellow); color(500pt)=(black); color(501pt)=(yellow); color(1000pt)=(orange); color(1500pt)=(red)},
	}

\usetikzlibrary{spy,backgrounds}
\usetikzlibrary{arrows,decorations.pathmorphing,backgrounds,positioning,fit,matrix}
\usepgfplotslibrary{external}
\fi

\usepackage{caption} 

\usepackage{mrm_submission}

\begin{document}


\graphicspath{{Figures/}}


\begin{center}
{\Large \bf \Papertitle
\vspace{6mm}

}
{\bf Jakob Assländer$^{1,2,3}$, Martijn A Cloos$^{1,2}$, Florian Knoll$^{1,2}$, Daniel K Sodickson$^{1,2}$, Jürgen Hennig$^{3}$ and Riccardo Lattanzi$^{1,2}$}

{$^1$
Bernard and Irene Schwartz Center for Biomedical Imaging, Department of Radiology, New York University School of Medicine, New York, NY, USA
}

{$^2$
Center for Advanced Imaging Innovation and Research (CAI$^2$R), Department of Radiology, New York University School of Medicine, New York, NY, USA
}

{$^3$
	Department of Radiology - Medical Physics, University Medical Center Freiburg, Germany
}
\vspace{4mm}

{\today}

\vspace{4mm}

Submitted to \textit{Magnetic Resonance in Medicine}

\end{center}

\vspace{1cm}

\noindent
Corresponding Author: \\
Dr. Jakob Assländer\\ 
Center for Biomedical Imaging\\
Department of Radiology\\
New York University School of Medicine\\
660 1st Avenue, New York, NY 10016, USA\\
Email: jakob.asslaender@nyumc.org
\vspace{1.5ex}


\vspace{0.5cm}

Word count: Abstract: 192/200, Body: 6067/5000

\newpage
\noindent
\section{Abstract}
\subsection{Purpose}
The proposed reconstruction framework addresses the reconstruction accuracy, noise propagation and computation time for Magnetic Resonance Fingerprinting (MRF).

\subsection{Methods}
Based on a singular value decomposition (SVD) of the signal evolution, MRF is formulated as a low rank inverse problem in which one image is reconstructed for each singular value under consideration. This low rank approximation of the signal evolution reduces the computational burden by reducing the number of Fourier transformations. Also, the low rank approximation improves the conditioning of the problem, which is further improved by extending the low rank inverse problem to an augmented Lagrangian that is solved by the alternating direction method of multipliers (ADMM). The root mean square error and the noise propagation are analyzed in simulations. For verification, in vivo examples are provided. 

\subsection{Results}
The proposed low rank ADMM approach shows a reduced root mean square error compared to the original fingerprinting reconstruction, to a low rank approximation alone and to an ADMM approach without a low rank approximation. Incorporating sensitivity encoding allows for further artifact reduction. 

\subsection{Conclusion}
The proposed reconstruction provides robust convergence, reduced computational burden and improved image quality compared to other MRF reconstruction approaches evaluated in this study.

\newpage
\section{Introduction}
The dynamics of large spin-1/2 ensembles in a magnetic field are generally described by the Bloch Equation, which captures spin-lattice and spin-spin interactions by relaxation terms with the characteristic time constants $T_1$ and $T_2$, respectively. Quantifying these parameters accurately, reproducibly and within a clinically feasible measurement time is desirable, e.g., for long term studies and for creating synthetic contrasts. 

Following departure from thermal equilibrium, otherwise undisturbed magnetization relaxes exponentially, which makes an inversion- or saturation-recovery experiment the natural choice for $T_1$-mapping. In order to keep the measurement time within limits, one usually acquires multiple data points during a single relaxation process \cite{Look1970}. The same goes for $T_2$-mapping, where a multi-spin echo sequence is usually employed. It has also been proposed to map $T_1$ and $T_2$ by acquiring multiple images with different flip angles, which proved to be very time efficient \cite{Crawley1988,Cheng2006,Deoni2003,Deoni2004}. However, these fast mapping techniques necessitate multiple interactions with the spin ensemble, which hampers the free relaxation and introduces systematic errors due to magnetization transfer and stimulated echoes amongst other effects. 

In recent years, model based parameter mapping has become increasingly popular for multiple reasons. Doneva et al. \cite{Doneva2010} used a pre-learned orthogonal dictionary to accelerate inversion-recovery and multi-spin echo based $T_1$ and $T_2$ mapping experiments. They proposed to reconstruct images from undersampled data by enforcing sparsity in the temporal domain, i.e., they assumed that few entries of the dictionary suffice for describing the signal evolution. Alternatively, model based parameter mapping can be used to remove systematic errors in fast imaging techniques. For example, model based $T_2$ mapping with multi-spin echo experiments has been shown to increase precision when incorporating signal contributions from stimulated echoes \cite{Sumpf2014, Ben-Eliezer2016}. 

The model based approach Magnetic Resonance Fingerprinting (MRF) \cite{Ma2013} aims at combining acceleration and accuracy. MRF departs from traditional sequence design concepts by deliberately avoiding a steady-state magnetization and relies on the Bloch equations to interpret the measured signal. Without the constraints imposed by a steady-state (or exponential relaxation), MRF opens up a rich new sequence design space that could fundamentally transform the way MR experiments are performed. For example, multiple transmit channels can be incorporated directly into the sequence to enable quantitative imaging with heterogeneous RF fields in high-field MR systems \cite{Cloos2016} and gradient waveforms can be modulated freely to transform the sound of a traditional MR systems into music \cite{Ma2016}.

In MRF, the excitation state is, in general, distinct at each $TR$. Thus, different repetitions cannot easily be combined, making Nyquist-sampled gradient encoding impossible at a clinically feasible spatial resolution and within a reasonable measurement time. The original MRF paper addresses this issue with a two step procedure: First, the undersampled k-space data are projected back into image space, resulting in a series of aliased images. Second, the time series of each voxel is matched to a precomputed dictionary. Under the premises of a vanishing correlation between the aliasing artifacts in the time series of each voxel and the dictionary, the atom (dictionary entry) with the highest correlation is determined by a brute force search within the dictionary of pre-simulated spin evolutions. The relaxation times are then extracted via a lookup table and the relative proton density is given by the correlation coefficient multiplied by a normalizing factor. 

Recent developments suggest that significant improvements in image quality can be gained by feeding the data model to the image reconstruction \cite{Davies2014, Zhao2016, Asslander2016a}. Due to the non-linear nature of the Bloch equation, the MRF reconstruction problem is generally non-convex. Therefore, global convergence of MRF reconstruction algorithms - including the proposed one - is usually not guaranteed and a good initial guess is essential. The present paper addresses this issue by a low rank (LR) approximation \cite{Petzschner2011, Zhao2015, Doneva2016, Cline2016, Zhao2016a, Zhao2015b} of the voxels' signal evolution, to which the alternating direction method of multipliers (ADMM) \cite{Zhao2016,Asslander2016a, Boyd2011} is applied. The proposed algorithm splits the MRF reconstruction problem into two sub-problems that are solved alternately: A linear inverse problem with data consistency at its core and a dictionary matching step. Describing the signal evolution with full rank, the inverse problem is under-determined in practical MRF cases. It has been suggested to address this by adding a low rank promoting regularization term \cite{Zhao2015b}. Here, we explicitly limit the search to a low rank space, which allows us to formulate an over-determined inverse problem. In the limit of a well-conditioned inverse problem, the first ADMM iteration solves the MRF reconstruction problem. In practice, however, the inverse problem is typically ill-conditioned even in the low rank approximation. In this case, the matched signal provides regularization to the data consistency term in consecutive ADMM iterations, compared to the mere data consistency term. We demonstrate the capabilities and convergence behavior of the proposed algorithm with simulations and experiments under practical scenarios.

\section{Theory}
In the following we derive a formalism that describes the entire MRF experiment in a single equation. This might seem a little cumbersome at first, but it facilitates the derivation of the proposed method.

\subsection{Dynamic MRI Signal Equation}
This section describes the signal formation assuming an arbitrary time series of images, which is described by a single vector $\mathbf{x} \in \mathbb{C}^{N T}$, where $N$ denotes the number of voxels and $T$ is the number of time-frames. The time series of images has the following structure:
\begin{equation}
\mathbf{x} = (x_{1,1}, \ldots, x_{N,1}, \ldots, x_{1,T}, \ldots, x_{N,T})' ,
\end{equation}
where $'$ denotes a vector transpose. We further allow for the k-space trajectory to vary between different $TR$s. For simplicity we assume the trajectories have the same length $K$ in each $TR$, which does not imply a loss of generality. The measured signal is denoted by the vector $\mathbf{S} \in \mathbb{C}^{K T}$. With the formalism of the non-uniform fast Fourier transformation (nuFFT), we describe the measurement process by the simple linear equation
\begin{equation}
\mathbf{S} = \mathbf{G} \cdot \mathbf{F} \cdot \mathbf{x} , 
\label{eq:Signal_MRI}
\end{equation}
where $\mathbf{F} \in \mathbb{C}^{N T \times N T}$ is a block-diagonal matrix containing $T$ blocks on the diagonal, each representing the same Fourier transformation along all considered spatial dimensions (see Eq.~\eqref{eq:F_structure} in the Appendix). Note that $\mathbf{F}$ can be implemented as an FFT of each time-frame in each spatial dimension. The gridding operator $\mathbf{G} \in \mathbb{C}^{K T \times N T}$ grids the Cartesian k-space data onto the non-Cartesian trajectory. Since gridding is a frame-by-frame operation, $\mathbf{G}$ is also a block-diagonal matrix. 

\subsection{MRF Reconstruction Problem}
A key element of MRF is modeling the signal with a precomputed dictionary, where each atom describes the signal evolution over time given a particular set of relaxation parameters (and potentially other effects). Thus, we can denote the dictionary by $\boldsymbol{\delta} \in \mathbb{C}^{T \times A}$, where $A$ is the number of atoms in the dictionary. After normalizing all atoms to have a unit $\ell_2$-norm, we can identify $T_1$ and $T_2$ of each voxel by finding the atom that maximizes the inner product with the estimated time series \cite{Ma2013}. The best fitting atoms of all voxels can be combined into a single matrix $\mathbf{D} \in \mathbb{C}^{NT \times N}$ which has the following structure:
\begin{equation}
\mathbf{D} = \begin{pmatrix}
d_{1,1}  & \cdots & 0          \\ 
\vdots & \ddots &  \vdots  \\
0          & \cdots & d_{N,1} \\
\vdots & \vdots & \vdots \\
d_{1,T} & \cdots & 0          \\ 
\vdots & \ddots &  \vdots  \\
0          & \cdots & d_{N,T} \\
\end{pmatrix} .
\label{eq:D}
\end{equation}
The non-zeros elements of each column in $\mathbf{D}$ correspond to a dictionary atom. The product $\mathbf{D}^H \mathbf{x}$ is a vector of scaling factors that reflect the proton density in each voxel, where $^H$ denotes the Hermitian conjugate (or conjugate transpose). Consequently, $\mathbf{D} \mathbf{D}^H \mathbf{x}$ is a time series of images composed by dictionary entries that are scaled to match $\mathbf{x}$. MRF generally assumes the signal evolution of all voxels to be described by the dictionary and the complete forward model is given by 
\begin{equation}
\mathbf{S} = \mathbf{G} \cdot \mathbf{F} \cdot \mathbf{D} \cdot \mathbf{D}^H \cdot \mathbf{x} .
\label{eq:Signal_MRF}
\end{equation}
The general MRF reconstruction problem can then be formulated as
\begin{equation}
\min_{\mathbf{D}, \mathbf{x}} \parallel \mathbf{G} \cdot \mathbf{F} \cdot \mathbf{D} \cdot \mathbf{D}^H \cdot \mathbf{x} - \mathbf{S} \parallel_2^2.
\label{eq:MRF_Problem}
\end{equation}
The two unknown $\mathbf{D}$ and $\mathbf{x}$ are multiplicative such that Eq. \eqref{eq:MRF_Problem} is a non-convex optimization problem. 

\subsection{Original MRF Reconstruction}
Ma et al. \cite{Ma2013} used a two-step procedure to solve Eq. \eqref{eq:MRF_Problem}. First, they used a filtered back-projection (BP) algorithm for reconstructing the time series of images. In the notation of Eq.~\eqref{eq:Signal_MRI}, such a reconstruction algorithm is given by
\begin{equation}
\mathbf{x}_{\text{BP}} = \mathbf{F}^H \cdot \mathbf{G}_{dc}^H \cdot \mathbf{S} , 
\label{eq:MRF_BP}
\end{equation}
where $\mathbf{G}_{dc}^H$ is the density compensated regridding operator. 

Given the time series of images $\mathbf{x}_{\text{BP}}$, the dictionary is used to determine the relaxation times for each voxel. While the original MRF reconstruction finds the best dictionary match using the maximum correlation, the following $\ell_2$-norm can equivalently be minimized:
\begin{equation}
\mathbf{D}_{\text{BP}} = \arg \min_{\mathbf{D}} \parallel \mathbf{x}_{\text{BP}} - \mathbf{D} \cdot \mathbf{D}^H \cdot \mathbf{x}_{\text{BP}} \parallel_2^2 .
\label{eq:l2_min_match}
\end{equation}
Here, each voxel (column of $\mathbf{D}$) can treated separately. By an exhaustive search in the dictionary, the atom is chosen that best matches the time series of the corresponding voxel in the least square sense. 

\subsection{Low Rank MRF Reconstruction Problem}
As shown in \cite{McGivney2014}, the dictionary can be compressed by a singular value decomposition:
\begin{equation}
\boldsymbol{\delta} = \mathbf{u} \boldsymbol{\Sigma} \mathbf{v}^H .
\label{eq:SVD}
\end{equation}
A low rank approximation of the dictionary matrix is given by
\begin{equation}
\tilde{\boldsymbol{\delta}} = \mathbf{u}_R^H \cdot \boldsymbol{\delta},
\end{equation}
where $\mathbf{u}_R \in \mathbb{C}^{T \times R}$ contains the first $R$ columns of the unitary matrix $\mathbf{u}$ and $R$ is the rank of the approximation. From here on, a tilde will denote variables associated with the low rank approximation.

Assuming that the time series of each voxel in $\mathbf{x}$ can be described by one atom of the dictionary, $\mathbf{x}$ can be approximated by 
\begin{equation}
\tilde{\mathbf{x}} = \mathbf{U}_R^H \cdot \mathbf{x}, 
\label{eq:compression}
\end{equation}
where $\mathbf{U}_R \in \mathbb{C}^{NT \times NR}$ is a block matrix composed of weighted identities:
\begin{equation}
\mathbf{U}_R = \begin{pmatrix}
u_{1,1} \cdot \mathbb{1} & \cdots & u_{1,R} \cdot \mathbb{1} \\
\vdots                                   & \ddots& \vdots \\
u_{T,1} \cdot \mathbb{1} & \cdots & u_{T,R} \cdot \mathbb{1} 
\end{pmatrix} .
\label{eq:UR}
\end{equation}
Here, $u_{t,r} \in \mathbb{C}$ are the entries of the first $R$ columns of the matrix $\mathbf{u}$ and $\mathbb{1}$ is the identity of size $N \times N$. The compressed dictionary matrix $\tilde{\mathbf{D}} \in \mathbb{C}^{NR \times N}$, can be calculated in the same way ($\tilde{\mathbf{D}} = \mathbf{U}_R^H \cdot \mathbf{D}$) and has the same structure as $\mathbf{D}$. Together with the low rank approximation of the image series $\tilde{\mathbf{x}} \in \mathbb{C}^{N R}$, the MRF reconstruction problem can be formulated as
\begin{equation}
\min_{\mathbf{\tilde{D}}, \tilde{\mathbf{x}}} \parallel \mathbf{G} \cdot \mathbf{F} \cdot \mathbf{U}_R \cdot \mathbf{\tilde{D}} \cdot \mathbf{\tilde{D}}^H \cdot \tilde{\mathbf{x}} - \mathbf{S} \parallel_2^2.
\end{equation}

A block diagonal matrix composed out of $R$ identical Fourier matrices is denoted by $\mathbf{\tilde{F}} \in \mathbb{C}^{NR \times NR}$ similar to $\mathbf{F} \in \mathbb{C}^{NT \times NT}$. The block diagonal structure of $\mathbf{\tilde{F}}$ and $\mathbf{F}$ in combination with $\mathbf{U}_R$ being composed of weighted identity matrices can be used to show that $\mathbf{F} \cdot \mathbf{U}_R = \mathbf{U}_R \cdot \mathbf{\tilde{F}}$ (see  Appendix). As mentioned earlier, one usually implements $\mathbf{\tilde{F}}$ as an FFT along all spatial dimensions. By changing the order of the Fourier transform and the SVD-compression, the computation time for the FFTs is reduced by a factor $R/T$. Further, $\mathbf{G}$ and $\mathbf{U}_R$ can be combined into the sparse matrix
\begin{equation}
\mathbf{\tilde{G}} \equiv \mathbf{G} \cdot \mathbf{U}_R
\end{equation}
prior to the actual reconstruction. This reduces the low rank MRF reconstruction problem to 
\begin{equation}
\min_{\mathbf{\tilde{D}}, \tilde{\mathbf{x}}} \parallel \mathbf{\tilde{G}} \cdot \mathbf{\tilde{F}} \cdot \mathbf{\tilde{D}} \cdot \mathbf{\tilde{D}}^H \cdot \tilde{\mathbf{x}} - \mathbf{S} \parallel_2^2.
\label{eq:MRF_LR_Problem}
\end{equation}
The evaluation of this objective function requires $R$ FFT operations (one for each considered singular value) and one sparse matrix-vector multiplication. The matrix $\mathbf{\tilde{G}}$ grids each of the resulting $R$ Cartesian k-space data sets onto the non-Cartesian k-space trajectory of the entire experiment. Therefore, the computational burden of the gridding operation is increased by a factor of $R$ compared to $\mathbf{G}$, which grids the k-space data frame by frame.

\subsection{Low Rank Back-Projection}
McGivney et al. \cite{McGivney2014} introduced the SVD compression for MRF in order to reduce the computation burden by matching $\tilde{\mathbf{x}}$ with the compressed dictionary. In the spirit of the original MRF reconstruction \cite{Ma2013}, they proposed to first reconstruct $\mathbf{x}_{BP}$ with Eq.~\eqref{eq:MRF_BP}, then compress the time series with Eq.~\eqref{eq:compression} and match the resulting $\tilde{\mathbf{x}}_{\text{BP}}$. This procedure is equivalent to the low rank back-projection
\begin{equation}
\tilde{\mathbf{x}}_{\text{BP}}= \mathbf{\tilde{F}}^H \cdot \mathbf{\tilde{G}}_{\text{dc}}^H \cdot \mathbf{S} .
\label{eq:SVD_BP}
\end{equation}
The matching procedure of Eq.~\eqref{eq:l2_min_match} is analogously given in the low rank case by
\begin{equation}
\mathbf{\tilde{D}}_{\text{BP}} = \arg \min_{\mathbf{\tilde{D}}} \parallel \tilde{\mathbf{x}}_{\text{BP}} - \mathbf{\tilde{D}} \cdot \mathbf{\tilde{D}}^H \cdot \tilde{\mathbf{x}}_{\text{BP}} \parallel_2^2 .
\label{eq:l2_min_match_svd}
\end{equation}

\subsection{Low Rank Inversion}
In general, a filtered back-projection, as used in Eq.~\eqref{eq:MRF_BP} and \eqref{eq:SVD_BP}, does not solve the given inverse problem. An inversion of the forward operation in \eqref{eq:MRF_BP} is not possible without additional constraints, since it is highly under-determined in practical MRF implementations ($KT \ll NT$)\footnote{Reminder: $K$ is the length of the trajectory of each time-frame, $N$ is the number of voxels of the image of each time-frame, $T$ is the number of time-frames and $R$ is the rank of the approximation.}. On the other hand, using the low rank approximation \eqref{eq:SVD_BP}, an overdetermined system ($KT \geq NR$) is obtained provided that $R \ll T$. In the derived formalism it is straightforward to formulate the inverse problem:
\begin{equation}
\tilde{\mathbf{x}}_{\text{inv}} = \arg \min_{\tilde{\mathbf{x}} \in \mathbb{C}^{N R}} \parallel \mathbf{\tilde{G}} \cdot \mathbf{\tilde{F}} \cdot \tilde{\mathbf{x}} - \mathbf{S} \parallel_2^2 .
\label{eq:SVD_Inversion}
\end{equation}
This is a linear system and can be solved with a conjugate gradient (CG) algorithm. This procedure searches for the low rank approximation $\tilde{\mathbf{x}}$ that best describes the measured signal in the least square sense. After solving Eq \eqref{eq:SVD_Inversion}, Eq. \eqref{eq:l2_min_match_svd} is used for the dictionary matching. 

\subsection{Low Rank Alternating Directions Method of Multipliers}
Depending on the particular implementation of the MRF experiment and the reconstruction, the condition $KT \geq NR$ might not be fulfilled. And even if it is, the condition number of the forward operation $\mathbf{\tilde{G}} \mathbf{\tilde{F}}$ with respect to inversion is likely high, making the above approach unfeasible. 

The recently proposed application of ADMM \cite{Boyd2011} to the original MRF \cite{Zhao2016,Asslander2016a} addresses the reconstruction problem by solving Eq. \eqref{eq:MRF_LR_Problem} via variable splitting. This approach can be denoted by an augmented Lagrangian:
\begin{equation}
\{\tilde{\mathbf{x}}_{\text{ADMM}}, \mathbf{\tilde{D}}_{\text{ADMM}}, \tilde{\mathbf{y}}_{\text{ADMM}}\} = \arg \min_{\tilde{\mathbf{x}}, \mathbf{\tilde{D}}, \tilde{\mathbf{y}}} \parallel \mathbf{\tilde{G}} \cdot \mathbf{\tilde{F}} \cdot \tilde{\mathbf{x}} - \mathbf{S} \parallel_2^2 + \mu \parallel \tilde{\mathbf{x}} - \mathbf{\tilde{D}} \mathbf{\tilde{D}}^H \tilde{\mathbf{x}} + \tilde{\mathbf{y}} \parallel_2^2 .
\label{eq:SVD_ADMM}
\end{equation}
The first summand is the data consistency term, which is equivalent to Eq.~\eqref{eq:SVD_Inversion}. The second summand compares the splitted variable $\tilde{\mathbf{x}}$ to its projection onto the dictionary $\mathbf{\tilde{D}} \mathbf{\tilde{D}}^H \tilde{\mathbf{x}}$. The Lagrangian multiplier $\tilde{\mathbf{y}} \in \mathbb{C}^{N R}$ is denoted here in the scaled dual form \cite{Boyd2011} and $\mu$ is the ADMM penalty parameter. 

The simultaneous search for the optimal $\tilde{\mathbf{x}}$, $\mathbf{\tilde{D}}$ and $\tilde{\mathbf{y}}$ in their product space is a challenging nonlinear problem. ADMM addresses this by alternately solving
\begin{align}
\tilde{\mathbf{x}}_{j+1} &= \arg \min_{\tilde{\mathbf{x}}} \parallel \mathbf{\tilde{G}} \cdot \mathbf{\tilde{F}} \cdot \tilde{\mathbf{x}} - \mathbf{S} \parallel_2^2 + \mu \parallel \tilde{\mathbf{x}} - \mathbf{\tilde{D}}_j \mathbf{\tilde{D}}_j^H \tilde{\mathbf{x}} + \tilde{\mathbf{y}}_j \parallel_2^2 
\label{eq:min_x}\\
\mathbf{\tilde{D}}_{j+1} &= \arg \min_{\mathbf{\tilde{D}}} \parallel \tilde{\mathbf{x}}_{j+1} - \mathbf{\tilde{D}} \mathbf{\tilde{D}}^H \tilde{\mathbf{x}}_{j+1} + \tilde{\mathbf{y}}_j \parallel_2^2 
\label{eq:min_D}\\
\tilde{\mathbf{y}}_{j+1} &= \tilde{\mathbf{y}}_{j} + \tilde{\mathbf{x}}_{j+1} - \mathbf{\tilde{D}}_{j+1} \mathbf{\tilde{D}}_{j+1}^H \tilde{\mathbf{x}}_{j+1} .
\label{eq:Lagrangian}
\end{align}
Eq.~\eqref{eq:min_x} is a linear optimization problem and can be solved with a CG algorithm. The minimization in Eq.~\eqref{eq:min_D} is solved by exhaustive search within the dictionary for each voxel. Note that Eq.~\eqref{eq:min_D} differs from the maximum correlation approach due to the Lagrangian multiplier, which increases the computational burden by a factor of two, whereas the complexity remains the same. The update of the Lagrangian multiplier (Eq.~\eqref{eq:Lagrangian}) has the standard form used for ADMM algorithms \cite{Boyd2011} and increasingly addresses those errors in $\tilde{\mathbf{x}}_{j} - \mathbf{\tilde{D}}_{j} \mathbf{\tilde{D}}_{j}^H \tilde{\mathbf{x}}_{j}$ that remain unchanged over multiple iterations. This avoids the necessity to continuously increase $\mu$ towards $\mu \rightarrow + \infty$ as $j$ increases in order to solve Eq.~\eqref{eq:MRF_LR_Problem}.

The optimization problem in Eq. \eqref{eq:SVD_ADMM} is non-convex, such that the convergence of the algorithm depends on the initial guess \cite{Zhao2016}. Here, we initialize the algorithm with $\tilde{\mathbf{x}}_0=0$, $(\mathbf{\tilde{D}}\mathbf{\tilde{D}}^H)_0 = \mathbb{1}$ and $\tilde{\mathbf{y}}_0=0$. Note that $(\mathbf{\tilde{D}}\mathbf{\tilde{D}}^H)_0 = \mathbb{1}$ is not described by the dictionary and was chosen to transform Eq. \eqref{eq:SVD_ADMM} into Eq. \eqref{eq:SVD_Inversion} for the first iteration. In the limit of a well conditioned $\mathbf{\tilde{G}} \mathbf{\tilde{F}}$ with respect to inversion, the first ADMM iteration solves the MRF reconstruction problem. With a slightly ill-conditioned Eq. \eqref{eq:SVD_Inversion}, we expect the first ADMM iteration to approximate the solution well enough to result in good convergence. This is not the case, where the data consistency term is highly ill-conditioned, including cases, where it is under-determined as used in \cite{Zhao2016,Asslander2016a}.

\subsection{Sensitivity Encoding}
Parallel imaging \cite{Sodickson1997} can help to further improve the conditioning of the reconstruction. Here, CG sensitivity encoding (SENSE) \cite{Pruessmann2001} is employed by reformulating the data consistency term:
\begin{equation}
\{\tilde{\mathbf{x}}_{\text{ADMM}}, \mathbf{\tilde{D}}_{\text{ADMM}}, \tilde{\mathbf{y}}_{\text{ADMM}}\} = \arg \min_{\tilde{\mathbf{x}}, \mathbf{\tilde{D}}, \tilde{\mathbf{y}}} \;\; \sum_{c} \parallel \mathbf{\tilde{G}} \cdot \mathbf{\tilde{F}} \cdot \mathbf{\tilde{E}}_c \cdot \tilde{\mathbf{x}} - \mathbf{S}_c \parallel_2^2 + \mu \parallel \tilde{\mathbf{x}} - \mathbf{\tilde{D}} \mathbf{\tilde{D}}^H \tilde{\mathbf{x}} + \tilde{\mathbf{y}} \parallel_2^2 .
\label{eq:SENSE}
\end{equation}
Before the Fourier transformation, the search vector $\tilde{\mathbf{x}}$ is multiplied by the diagonal matrix $\mathbf{\tilde{E}}_c \in \mathbb{C}^{NR \times NR}$ which contains $R$ repetitions of the sensitivity profile of coil $c$ and is then compared to the signal $\mathbf{S}_c$ that was measured by the coil under consideration. The sum over all coil elements provides the total cost of the data-consistency term. Note that SENSE can also be incorporated in the low rank inversion (Eq. \eqref{eq:SVD_Inversion}) in the same way. 

\newpage
\section{Methods}
In the spirit of reproducible research, the source code of the proposed algorithm is available at \url{https://bitbucket.org/asslaender/nyu_mrf_recon}.

\subsection{Simulations}
All simulations were performed with the pseudo steady-state free precession (pSSFP) pattern described in \cite{Asslander2016a}. The flip angle scheme is displayed in Fig.~\ref{fig:TimeSeries}a and consists of an inversion pulse followed by RF-pulses with varying flip angles and a phase increment of $\pi$ between consecutive pulses. The repetition and echo time of each repetition was calculated based on the pSSFP approach with $TR_0 = 5$~ms. A dictionary was computed with Bloch simulations with $T_1 (\text{s}) = 0.3 \cdot 1.02^j \; \forall \; j \in \{0, 1, \ldots, 152\}$, thus covering the range between $300$~ms and $6$~s in steps of $2$\%. The dictionary covers $T_2$ between $50$~ms and $3$~s in steps of $2$\%, i.e. $T_2 (\text{s}) = 0.05 \cdot 1.02^j \; \forall \; j \in \{0, 1, \ldots, 207\}$. For simplicity, off-resonance effects were neglected, as well as $B_1$ inhomogeneities and other interfering factors such as slice profile imperfections and magnetization transfer. The resulting dictionary $\boldsymbol{\delta} \in \mathbb{R}^{841 \times 24\;921}$ has a rank of $62$ at a tolerance of $2.8 \cdot 10^{-9}$. The dictionary is real valued due to the nature of the sequences and the assumption of on-resonant spins. An SVD was performed according to Eq.~\eqref{eq:SVD}. The first five left-singular vectors (columns of $\mathbf{u}$) are depicted in Fig.~\ref{fig:TimeSeries}b and the corresponding singular values in (c).

A numerical brain phantom from the Brainweb database (\url{http://brainweb.bic.mni.mcgill.ca/brainweb/}) \cite{Collins1998} was used to create synthetic data sets. Segments of fat were associated with the "original" values $PD^{\text{org}}=0.9$, $T_1^{\text{org}} = 0.37$~s, $T_2^{\text{org}} = 0.13$~s, white matter (WM) with  $PD^{\text{org}}=0.65$, $T_1^{\text{org}} = 1.08$~s, $T_2^{\text{org}} = 0.07$~s, gray matter (GM) with $PD^{\text{org}}=0.8$, $T_1^{\text{org}} = 1.82$~s, $T_2^{\text{org}} = 0.10$~s and cerebrospinal fluid (CSF) with $PD^{\text{org}}=1.0$, $T_1^{\text{org}} = 4.5$~s, $T_2^{\text{org}} = 2.2$~s. Bloch simulations were used to create the original time series $\mathbf{x}_\text{org}$. Unless stated otherwise, a matrix size of $128 \times 128$ was used, which corresponds to $N = 16384$. If not stated otherwise, one radial k-space spoke with a golden angle increment \cite{Winkelmann2007} was acquired for each $TR$. To demonstrate the impact of sampling density on the reconstruction quality, a subset of the simulations was repeated with multiple spokes per TR. These additional spokes where distributed using the tiny golden angle of approximately $32.0397^\circ$ \cite{Wundrak2015}. All radial spokes were associated with an extend of $k_{\max} = \sqrt{2} \pi/\text{voxel}$ contrary to the usual convention of radial MRI ($k_{\max} = \pi/\text{voxel}$). This avoids artifacts due to unsampled corners of k-space in the simulation. The signal $\mathbf{S}$ was simulated according to Eq.~\eqref{eq:Signal_MRI}, where the forward operation $\mathbf{GF}$ was implemented as a nuFFT operator with Kaiser-Bessel interpolation and optimal scaling in the min-max sense \cite{Fessler2003}. Three nearest neighbors were taken into account in each spatial direction and an oversampling factor of 2 was used. 

\subsubsection{Image Reconstruction}
Four different reconstructions were performed on the same noise-free synthetic data: A filtered back-projection (Eq.~\eqref{eq:MRF_BP}) as proposed in \cite{Ma2013}, a filtered back-projection in the low rank domain (Eq.~\eqref{eq:SVD_BP}) as proposed in \cite{McGivney2014}, a low rank approximation of the signal (Eq. \eqref{eq:SVD_Inversion}) and the proposed low rank ADMM reconstruction (Eq. \eqref{eq:SVD_ADMM}). The back-projection reconstructions were performed by first gridding the data onto an enlarged k-space (factor 2) with linear density compensation. Thereafter, k-space was cut to cover $\pm \pi/\text{voxel}$ followed by an FFT. This procedure ensures correct density compensation with the lengthened radial spokes. All CG-based methods do not rely on density compensation such that this procedure can be avoided. The low rank inversion was performed for $R \in \{2,3,\ldots,6,8,\ldots,20\}$. Eq.~\eqref{eq:SVD_Inversion} is solved by a CG algorithm with 100 iterations after initializing it with zero. The low rank ADMM reconstructions were performed with 20 CG steps for solving Eq.~\eqref{eq:min_x} in each ADMM iteration. A simulation was performed with $R \in \{2,3,\ldots,6,8,\ldots,20\}$ at a fixed $\mu = 1.26 \cdot 10^{-3}$ and 10 ADMM iterations. Further, at a fixed $R=5$, reconstructions were performed for the product space of $\mu = 10^{j} \; \forall \; j \in \{-4, -3.8, \ldots, -2\}$ and the number of ADMM iterations $\in \{1,2,\ldots, 31\}$. 

\subsubsection{Root Mean Square Error}
The root mean square error (RMSE) was calculated from noise-free reconstructions. A region of interest was drawn over the entire white matter and the reconstructed quantitative maps were compared to the original maps of $PD^{\text{org}}, T_1^{\text{org}}$ and $T_2^{\text{org}}$. The RMSE was normalized by the original value and is then referred to as NRMSE. 

\subsubsection{Noise Analysis}
The reconstructions with $R=5$, $\mu = 1.26 \cdot 10^{-3}$, 10 ADMM and 20 CG iterations were repeated 100 times each with pseudo random white noise added to the signal $\mathbf{S}$ \cite{Robson2008}. The input noise level was $SNR_{\text{in}} = 100$ for each case, where $SNR_{\text{in}} = 1$ corresponds an equal mean and standard deviation in a voxel with $PD^{\text{org}} = 1$ in a hypothetical experiment with $\alpha = 90^\circ$, $TR \rightarrow + \infty$ and $TE \rightarrow +0$. 

Two different analyses were performed on the $100$ pseudo replicas: First, the $PDNR$, $T_1NR$ and $T_2NR$ were calculated by dividing the mean $PD$, $T_1$ and $T_2$ of all white matter voxels in all pseudo replicas by the corresponding standard deviation. This analysis was performed for the low rank inversion and the low rank ADMM reconstruction and the result was normalized by $SNR_{\text{in}}$.

A second analysis was performed on the pseudo replicas reconstructed with a rank $R=5$. For each tissue, the mean and the standard deviation of all voxels and all pseudo replicas were calculated. They were both normalized by the original parameters used in the simulation.

\subsubsection{Sensitivity Encoding}
The low rank ADMM simulation (Eq. \eqref{eq:SVD_ADMM}) was repeated with an increased matrix size of $384 \times 384$ at the same number of spokes in order to increase undersampling artifacts. It is compared to a reconstruction that incorporates SENSE (Eq. \eqref{eq:SENSE}). In vivo coil sensitivity profiles of 12 virtual coils of a 44 channel head coil (Siemens, Erlangen, Germany) were employed. The numerical phantom was multiplied by the sensitivity profile of each coil element. Thereafter, the signal of each coil was simulated with Eq. \eqref{eq:Signal_MRI} followed by a reconstruction based on Eq. \eqref{eq:SENSE} with the same parameters as used before.

\subsection{In Vivo Experiments}
Single slice MR fingerprinting data of a healthy volunteer’s brain were acquired with a 3T Skyra scanner (Siemens, Erlangen, Germany). Written informed consent was obtained and the protocol was approved by the Institutional Review Board. For signal reception, the manufacturer’s 16-channel head coil was used, which were reduced to 8 virtual coils via SVD compression. The previously described pSSFP acquisition scheme was employed. The inversion pulse was implemented as a $10.24$~ms secant pulse followed by a train of sinc-pulses with  a time bandwidth product of 2 and a duration of $800~\upmu \text{s}$. The in plane spatial resolution was $1~\text{mm} \times 1~\text{mm}$ at a $FOV = 192~\text{mm} \times 192~\text{mm}$ with a slice thickness of 3~mm. The dwell time of the readout was set to $2.4~\upmu \text{s}$. With these parameters, the timing of the $i^\text{th}$ readout can be realized only when $\min \{TE_i, TR_i - TE_i \} > 1.3$~ms. The current implementation of the sequence skips the readout whenever this requirement is not fulfilled. In total, 841 of the designated 850 spokes were acquired within $4.1~\text{s}$. After acquiring the data along one RF-train, a $10~\text{s}$ pause was inserted to ensure full relaxation before the next repetition of the RF-train is started. Relative to one another, subsequent spokes within the RF-train are rotated by the golden angle ($111.25^\circ$) \cite{Winkelmann2007}. Each time the RF-train is repeated, all spokes are rotated by the tiny golden angle of approximately $32.0397^\circ$ \cite{Wundrak2015}. A total of 32 repetitions were measured. The data set was reconstructed in two different ways, using all samples and using only the first RF-train. For more details about the acquisition, please refer to \cite{Asslander2016a}, where a similar experimental setup was used.

For image reconstruction, the nuFFT was implemented with Kaiser-Bessel interpolation and optimal scaling in the min-max sense \cite{Fessler2003}. Three nearest neighbors were considered in each direction and an oversampling of 2 was employed. The low rank inversion was performed with 20 CG steps and the low rank ADMM reconstruction was performed with 5 ADMM iterations and each with 10 CG iterations. The ADMM penalty parameter was set to $\mu = 0.1$ based on the visual impression of the results.

\newpage
\section{Results}
\subsection{Simulations}
As shown in Fig.~\ref{fig:TimeSeries}c, the singular values of the dictionary matrix $\boldsymbol{\delta}$ decrease rapidly. Reconstructions were performed in low rank approximations with $R \in \{2, 3, \ldots, 6, 8, \ldots, 20\}$. The corresponding NRMSE based on noise free reconstructions are shown in Fig.~\ref{fig:RMSE_SNR}a-c. The signal to noise ratio based on pseudo replicas is displayed in (d-f). Using the low rank back-projection reconstruction (Eq.~\eqref{eq:SVD_BP} and \eqref{eq:l2_min_match_svd}), the NRMSE and the SNR are hardly influenced by the rank of the approximation (Fig.~\ref{fig:RMSE_SNR}a,d). However, solving the inverse problem in the low rank approximation (Eq.~\eqref{eq:SVD_Inversion}) reduces the overall NRMSE (b). The minimum NRMSE is reached at $R=3$, where also a local maximum of the SNR can be found, which is greater than the SNR of the back-projection reconstruction. At higher rank approximations, a non-monotonic behavior can be observed: The NRMSE of all parameters increases for $3 < R \leq 11$ and for $R>11$ the NRMSE of $T_2$ and $PD$ decreases again. Also, the SNR shows an non-monotonic behavior, which favors proton density over the relaxation times at $R>11$. Applying the ADMM algorithm to the low rank approximation, a further decreased NRMSE can be observed compared to the low rank inverse problem. Also, the minimum of the NRMSE is shifted towards $R=5$ and the SNR shows a more stable behavior with a maximum at $R=5$.

Fig. \ref{fig:RMSE_spokes}a repeats the results of Fig. \ref{fig:RMSE_SNR}c with a different scaling and compares them to the same simulation with 4 radial spokes per $TR$ instead of one. Since all other parameters were kept equivalent, this effectively reduces the undersampling. As a consequence, the NRMSE is reduced over all and the minimum is shifted to $R=8$. Note that the NRMSE of $T_1$ is bound by the discretization of the dictionary in this particular example. 

\subsubsection{Reconstructions with $R=5$}
Based on the NRMSE minimum, $R=5$ was selected for a more detailed analysis of the data with one radial spoke per $TR$. The NRMSE and the residuum of the objective function are displayed in Fig. \ref{fig:Convergence} as a function of the ADMM penalty parameter $\mu$ and the number of ADMM iterations. After 1 iteration (dark blue), $\mu$ has no influence on the NRMSE and the residuum, which is expected due to the initialization of the algorithm. With an increasing number of iterations, a monotonous decrease of the NRMSE and the residuum can be observed until the point of convergence, after which the NRMSE increases slightly in some cases. In general, the reduction speed of the NRMSE depends on $\mu$, as does the minimum NRMSE reached after convergence. The minimal NRMSE varies slighty as a function of $\mu$. For bigger $\mu$ values, the NRMSE abruptly increases. The optimal $\mu$ also depends on the number of iterations as well as the considered parameter ($PD, T_1, T_2$). For practical reasons, 10 iterations (black) were chosen here heuristically, for which $\mu = 1.26 \cdot 10^{-3}$ (arrows in Fig. \ref{fig:Convergence}) minimizes the NRMSE of $T_1$, while $PD$ and $T_2$ are close to optimal, justifying its choice for reconstructions throughout the paper. The residuum (Eq. \eqref{eq:MRF_LR_Problem}) converges very fast and shows only little variations within a certain range of $\mu$ and the minimum at 10 iterations is shifted with respect to the NRMSE (d).  

Fig.~\ref{fig:SVDRecon} shows the five elements of $\tilde{\mathbf{x}}$ reconstructed from noise free data with $\mu = 1.26 \cdot 10^{-3}$. The rows correspond to the singular values in decreasing order (the color coding corresponds to Fig. \ref{fig:TimeSeries}b,c). The right hand side of all images are displayed using the same scale and the decreasing brightness reflects the magnitude of the singular values. The second column depicts the back-projection reconstruction (Eq.~\eqref{eq:SVD_BP}). Contrary to the other reconstructions, the individual images do not depend on the rank of the approximation. The images are increasingly dominated by noise-like artifacts for decreasing singular values. The third column displays $\tilde{\mathbf{x}}_{inv}$, which was reconstructed with Eq.~\eqref{eq:SVD_Inversion} and exhibits considerably reduced artifacts. However, residual undersampling artifacts can be observed when increasing the brightness (left half of the images), especially for smaller singular values (last two rows). The artifacts are further reduced, when employing the proposed ADMM reconstruction (Eq.~\eqref{eq:SVD_ADMM}).

The corresponding quantitative maps are shown in Fig.~\ref{fig:Maps_nSVD_5}. They were reconstructed with Eq.~\eqref{eq:l2_min_match} (second column), \eqref{eq:l2_min_match_svd} (column 3 and 4) and \eqref{eq:SVD_ADMM} (column 5). The back-projection reconstruction (column 2) exhibits significant undersampling artifacts in all three parameter maps. Noise-like artifacts can be seen in the proton density, as well as in the $T_1$ map and most prominently  in the $T_2$ map. The quantitative maps reconstructed from the low rank back-projection (third column of Fig.~\ref{fig:Maps_nSVD_5}) and the time domain back-projection (second column) are visually the same. In fact, the low rank back-projection does not promise improved image quality and is purely motivated by computation time \cite{McGivney2014}. On the contrary, solving the inverse problem in the low rank approximation (Eq.~\eqref{eq:SVD_Inversion}) results in significantly improved quantitative maps: The noise-like artifacts are reduced, which is even more the case when employing the ADMM algorithm in the low rank domain. 

When analyzing the reconstruction of noisy data within all four simulated tissue types, systematic errors can be observed in the two cases of the back-projection reconstruction, most prominently for $T_2$ (Fig. \ref{fig:SNR}). On the contrary, both the low rank inversion and the low rank ADMM reconstruction accurately reproduce the original parameters, except in the case of CSF, where systemic errors in $T_1$ and $T_2$ can still be noticed. The employed excitation pattern in combination with the short acquisition time ($4.1~\text{s}$) results in a reduced sensitivity to long relaxation times. As a consequence, the relaxation times of CSF are not measured precisely. Furthermore, the standard deviation is decreased in all tissue types when employing the low rank ADMM reconstruction. Due to the non-linearity of the Bloch equation, the noise is distributed in a non-Gaussian manner (not shown here). Therefore, mean and standard deviation have a limited information value. For the cases with strong deviations from a Gaussian distribution, the standard deviation is not displayed.

\subsubsection{Reconstructions with $R=20$}
Fig. \ref{fig:Maps_nSVD_20} compares the low rank inversion and the ADMM algorithm for $R=20$ in order to further analyze the influence of the rank. In the case of the low rank inverse problem (Eq. \eqref{eq:SVD_Inversion}), spatial blurring can be observed, most prominently in the $T_1$ map. This is also reflected in Fig. \ref{fig:RMSE_SNR}b, where the NRMSE increase at high $R$ values is strongest for $T_1$. When employing the ADMM algorithm, some noise-like artifacts can be observed, but the blurring is reduced compared to the low rank inversion.

\subsubsection{Sensitivity Encoding}
The benefits of SENSE in the low rank ADMM reconstruction is demonstrated in Fig.~\ref{fig:SENSE}. Without parallel imaging, the ADMM reconstruction of higher resolution data with same number of radial spokes results in increased undersampling artifacts. These artifacts are reduced by incorporating CG SENSE \cite{Pruessmann2001} in the proposed low rank ADMM reconstruction.

\subsection{In Vivo Experiment}
The in vivo maps reconstructed from one radial spoke per $TR$ (Fig. \ref{fig:InVivo}) appear noisy when reconstructed with the standard back-projection and the low rank back-projection. However, the artifact level is reduced compared to the simulations. Since the data were acquired with a multi-coil array, the back-projection includes a multiplication with the complex conjugate of the coil sensitivity profiles. Even though this does not represent a parallel imaging algorithm, it still mitigates the undersampling artifacts to some extent. An improvement of the reconstruction quality can be observed when employing a low rank inversion and particularly when using the proposed low rank ADMM reconstruction.  However, the noise level is still high, especially in the $T_2$ map. This can be attributed to the short acquisition time ($4.1~\text{s}$) in combination with a suboptimal acquisition pattern. A similar effect can also be observed in the simulation data shown in Figs.~\ref{fig:RMSE_SNR}f and \ref{fig:SNR}, which also show a relatively high variability in the $T_2$ maps. Measuring the mean and standard deviation over a region of interest in the frontal white matter shows a decrease in the standard deviation when employing the proposed LR ADMM algorithm (Tab.~\ref{tab:in_vivo}). Note that the mean values are affected by slice profile imperfections and magnetization transfer effects.

When using 32 spokes per TR, all reconstruction techniques converge to very similar results (Fig.~\ref{fig:InVivo_32spokes}). This is also reflected by the comparably constant relaxation times measured within a region of interest in the frontal white matter (Tab.~\ref{tab:in_vivo}). Note that $B_0$ drifted significantly during the time needed to obtain all $32$ spokes. Consequently, the relaxation times are biased and deviate from those obtained when reconstructed from the first repetition only (Fig.~\ref{fig:InVivo}). This effect originates from the sensitivity of balanced SSFP sequences to $B_0$ variations and a detailed study of these effects is beyond the scope of this paper. 

\newpage
\section{Discussion}
In this paper we propose a reconstruction framework for MRF based on a low rank approximation of the signal evolution in combination with the alternating direction method of multipliers. The original MRF reconstruction \cite{Ma2013} and SVD-based dictionary matching \cite{McGivney2014} are described within a formalism that identifies SVD-based matching as a filtered back-projection into the space of few singular vectors. By exploiting the commutative property of the spatial Fourier transformation and the SVD-compression in the temporal domain (see Appendix), we show that the computation time of the FFTs can be reduced by a factor $R/T$, which is $5/841$ in the particular case under investigation. The gridding operator and the SVD-compression are combined into a single sparse matrix, which allows for gridding the Cartesian k-space data in the domain of the singular vectors directly onto the non-Cartesian trajectory in the temporal domain. The overall computational cost of this sparse matrix-vector multiplication is a factor of $R$ larger than the frame-by-frame gridding in the standard reconstruction, since $\mathbf{\tilde{G}}$ grids each singular value image onto the entire trajectory. However, an overall speedup is expected when setting $R \ll T$. For the in vivo data, a speedup of a factor of 12 was observed for executing $\mathbf{\tilde{F}}^H \mathbf{\tilde{G}}^H \mathbf{\tilde{G}} \mathbf{\tilde{F}} \tilde{\mathbf{x}}$ compared to $\mathbf{F}^H \mathbf{G}^H \mathbf{G} \mathbf{F} \mathbf{x}$. However, the speedup depends on $R$, $T$ and the implementation of the nuFFT. For large $T$ values as used, e.g., for 3D fingerprinting, even larger speedups are expected. Solving Eq. \eqref{eq:min_D} has roughly double the computational cost compared to maximizing the correlation, since $\mathbf{\tilde{D}}^H \tilde{\mathbf{x}}$ and $\mathbf{\tilde{D}}^H \tilde{\mathbf{y}}$ need to be calculated. However, the low rank approximation reduces the complexity of these matrix-vector multiplications and speeds up the matching process by the same factor as in the low rank back-projection proposed in \cite{McGivney2014}. The reconstruction of the in vivo data (Fig.~\ref{fig:InVivo}) took approximately $43~\text{s}$ to reconstruct on a desktop computer (3.3 GHz Intel i5-6600 processor). The reconstruction is implemented in MATLAB (The MathWorks, Natick, MA, USA). Further speedups could be achieved with a multi core or GPU sparse matrix multiplication and a more efficient dictionary search \cite{Cline2016}. 


The formalism introduced here recasts the reconstruction of the image series into the form of a straightforward inverse problem that is over-determined depending on the choice of parameters. When the condition $KT \geq NR$ is fulfilled, artifact free images can theoretically be reconstructed with Eq. \eqref{eq:SVD_Inversion}. Our low rank inversion embodies some concepts akin to the methods described in \cite{Petzschner2011} in the context of parameter mapping and more specifically in \cite{Doneva2016,Zhao2016a} in the context of MRF. However, the method described in \cite{Zhao2016a} does not provide the here described speed-up related the number FFT operations, while the method proposed in \cite{Doneva2016} lacks a straightforward incorporation of sensitivity encoding. Fig. \ref{fig:RMSE_SNR} shows that very small $R$ values result in approximation errors. When $R$ is increased, the problem first becomes increasingly ill-conditioned, which results in strong noise amplification (Fig. \ref{fig:RMSE_SNR}). If the rank is increased beyond $R=11$, the problem becomes under-determined ($KT < NR$). Employing a CG algorithm to this under-determined problem leads to blurring, particularly in the $T_1$ map (Fig. \ref{fig:Maps_nSVD_20}). On the other hand, the SNR of $T_1$ decreases for $R>11$ (Fig. \ref{fig:RMSE_SNR}). This identifies the blurring not to be a simple smoothing effect, where an increased SNR for $T_1$ would be expected. 

The described issues can be mitigated by applying the proposed ADMM algorithm to the low rank approximation of the time series. ADMM, like other iterative reconstructions proposed for MRF, constrains the reconstruction directly to atoms of the dictionary \cite{Davies2014,Zhao2016,Asslander2016a,Zhao2015b}. However, these approaches are non-convex problems which are prone to convergence issues. Examining Fig. \ref{fig:RMSE_SNR}c and f, an increased NRMSE and a decreased SNR can be observed for an increasing $R$. This shows that the ADMM algorithm without a low rank approximation converges to a solution that neither minimizes NRMSE nor maximizes SNR. Note that a reconstruction with $R = 841$ would be equivalent to an ADMM reconstruction in the temporal domain without a low rank approximation similar to \cite{Zhao2016,Asslander2016a}. The application of the ADMM algorithm to $\tilde{\mathbf{x}}$ in the space of the first $R$ singular vectors reduces the NRMSE and increases the SNR compared to a low rank approximation alone (Fig. \ref{fig:RMSE_SNR} b vs. c and e vs. f) and also compared to the ADMM algorithm without a low rank approximation (Fig. \ref{fig:RMSE_SNR} c and f: low $R$ vs. high $R$). In general, choosing the rank of the approximation is a trade-off between not describing the signal evolution accurately enough (small $R$) and solving an ill-posed problem (large $R$). The first part of the trade-off depends on the dictionary and ultimately on the RF-, $TR$- and $TE$-train, as well as the types of tissue under consideration, while the latter part depends on the spatial encoding (Fig. \ref{fig:RMSE_spokes}). Consequently, the best choice of $R$ depends on the pulse sequence. However, Fig. \ref{fig:RMSE_SNR} indicates a reduced sensitivity of the ADMM approach to large $R$ values compared to the low rank inversion alone, which is expected due to extra penalty term that is added in Eq. \eqref{eq:SVD_ADMM} compared to Eq. \eqref{eq:SVD_Inversion}. 

In the case of convex optimization, the ADMM penalty parameter $\mu$ influences the speed of convergence, but not the solution \cite{Boyd2011}. This is not the case for non-convex optimization problems as Eq. \eqref{eq:SVD_ADMM}, where $\mu$ can also influence the solution after convergence (Fig. \ref{fig:Convergence}). The extreme values give some intuition about the effect of $\mu$. The ADMM reconstruction problem (Eq. \eqref{eq:SVD_ADMM}) approaches the low rank inversion (Eq.~\eqref{eq:SVD_Inversion}) for small $\mu$, while large $\mu$ emphasize matching $\tilde{\mathbf{x}}$ to $\mathbf{\tilde{D}}_j \mathbf{\tilde{D}}_j^H \tilde{\mathbf{x}}$ after the first iteration. As a consequence, data consistency is neglected and the resulting parameter maps exhibit little change after the initial guess. This is likely the cause of the abrupt increase of the NRMSE when increasing $\mu$ beyond a certain point. This abrupt increase suggests to choose smaller $\mu$ in practical cases. If the algorithm is terminated before convergence, the optimal $\mu$ is slightly shifted (black dots and arrows in Fig. \ref{fig:Convergence}). The minimal residuum does in general not coincide with the smallest NRMSE, which limits the feasibility of using it to tune $\mu$. For the in vivo data, $\mu$ was chosen heuristically. The step size of the relaxation times in the dictionary does generally influence convergence as well. A finer grid allows $\mathbf{\tilde{D}}$ to be updated even after small changes in $\tilde{\mathbf{x}}$, potentially improving the convergence. However, the present implementation penalizes the distance to the dictionary, rather than strictly tying $\tilde{\mathbf{x}}$ to the dictionary, as done in the BLIP algorithm \cite{Davies2014}. Therefore, the step size in the dictionary is not expected to be crucial for convergence and a detailed analysis is omitted here. 

Due to the non-linearity of the Bloch equation and the reconstruction, the noise in the computed quantitative maps is not expected to be Gaussian distributed. A general reduction of the noise can be observed for the proposed low rank ADMM approach compared to the other reconstructions under investigation (Fig. \ref{fig:SNR}). In the case of CSF, $T_1$ and $T_2$ exhibit noise-like variations that deviate strongly from a Gaussian distribution (not shown here). However, the SNR and the distribution of the noise depends strongly on the acquisition pattern and a careful consideration of these effects is beyond the scope of this paper.

Formulating the reconstruction as a minimization problem (Eq. \eqref{eq:SVD_ADMM}) provides the flexibility to add different constraints. Here, iterative SENSE was incorporated (Eq. \eqref{eq:SENSE}, Fig. \ref{fig:SENSE}, \ref{fig:InVivo} and \ref{fig:InVivo_32spokes}). Additional priors in the spirit of compressed sensing \cite{Lustig2007} can also be added as additional summands to Eq. \eqref{eq:SVD_ADMM}. The provided source code implements an $\ell_{1,2}$-penalty of $\tilde{\mathbf{x}}$ \cite{Zhao2015} (not shown here). Future work could also explore the potential advantages of incorporating the nuclear norm which has been shown to provide superior properties for multi-contrast image reconstruction \cite{Knoll2016}. 

\section{Conclusion}
A reconstruction framework for MRF is presented that incorporates the data model in two different ways. A low rank approximation of the signal evolution is computed based on a singular value decomposition of the dictionary matrix. Additionally, the best matching atom of the dictionary is used to constrain the signal evolution in each voxel. An ADMM algorithm is then used to alternately optimize the low rank approximation of the signal evolution and its projection onto the dictionary. The resulting quantitative maps show a reduced error and superior SNR compared to the original MRF reconstruction, a low rank approximation alone, as well as the application of the ADMM algorithm in the temporal domain rather than in a low rank space. The proposed algorithm is a general reconstruction framework for MRF and can be readily applied to plug and play parallel transmission \cite{Cloos2016} and chemical exchange MRF-X \cite{Hamilton2015} amongst others.

\section{Acknowledgments}
This work was supported in part by the research grants NIH R21 EB020096 and was performed under the rubric of the Center for Advanced Imaging Innovation and Research (CAI2R, www.cai2r.net), a NIBIB Biomedical Technology Resource Center \newline (NIH P41 EB017183).

\appendix
\section{Appendix}
\subsection{Commutation of the Fourier transformation and the compression matrix}
In order to prove $\mathbf{\tilde{F}} \cdot \mathbf{U}_R = \mathbf{U}_R \cdot \mathbf{F} $, we denote $\mathbf{F}$ as a $T \times T$ block matrix containing the Fourier matrix $\mathbf{f} \in \mathbb{C}^{N \times N}$ on the diagonal and zero-matrices elsewhere: 
\begin{equation}
\mathbf{F} = \begin{pmatrix}
\mathbf{f} & \mathbf{0} & \cdots         & \mathbf{0} \\
\mathbf{0} & \mathbf{f} & \ddots         & \mathbf{0} \\
\vdots         & \ddots        & \ddots         & \mathbf{0} \\
\mathbf{0} & \cdots        & \mathbf{0} & \mathbf{f} \\
\end{pmatrix} .
\label{eq:F_structure}
\end{equation}
In combination with the notation in Eq.~\eqref{eq:UR}, the product of $\mathbf{U}_R$ and $\mathbf{F}$ can be written as
\begin{equation}
\mathbf{U}_R \cdot \mathbf{F} = \begin{pmatrix}
u_{1,1} \cdot \mathbb{1} \cdot \mathbf{f} & \cdots & u_{T,1} \cdot \mathbb{1} \cdot \mathbf{f}  \\
\vdots                                   & \ddots& \vdots \\
u_{1,R}  \cdot \mathbb{1} \cdot \mathbf{f} & \cdots & u_{T,R}  \cdot \mathbb{1}  \cdot \mathbf{f}
\end{pmatrix} .
\label{eq:URF}
\end{equation}
If we denote $\mathbf{\tilde{F}}$ as an $R \times R$ block-diagonal matrix with the same matrix $\mathbf{f}$ on the diagonal, we can equivalently denote
\begin{equation}
\mathbf{\tilde{F}} \cdot \mathbf{U}_R = \begin{pmatrix}
\mathbf{f} \cdot u_{1,1} \cdot \mathbb{1} & \cdots & \mathbf{f} \cdot u_{T,1} \cdot \mathbb{1}  \\
\vdots                                   & \ddots& \vdots \\
\mathbf{f} \cdot u_{1,R}  \cdot \mathbb{1} & \cdots & \mathbf{f} \cdot u_{T,R}  \cdot \mathbb{1}
\end{pmatrix} .
\label{eq:FtUR}
\end{equation}
Since the scalars $u_{t,r}$ and the unit matrix $\mathbb{1}$ always commute, Eq.~\eqref{eq:URF} and \eqref{eq:FtUR} can be combined to yield
\begin{equation}
\mathbf{\tilde{F}} \cdot \mathbf{U}_R = \mathbf{U}_R \cdot \mathbf{F} .
\end{equation}

\bibliographystyle{mrm}
\bibliography{library}

\section{Figure Captions}

\subsection*{Figure \ref{fig:TimeSeries}}
\def\CaptTimeSeries{
	The flip angle pattern is depicted in (\textbf{a}), while (\textbf{b}) shows the first five left singular vectors (columns of $u$), which result from the singular value decomposition of the dictionary and were used for the low rank approximation unless stated otherwise. Their color-coding corresponds to (\textbf{c}), where the first ten singular values of the dictionary are shown. 
}
\CaptTimeSeries

\subsection*{Figure \ref{fig:RMSE_SNR}}
\def\CaptRMSESNR{
	The normalized root mean square error (NRMSE) in white matter is shown (\textbf{a}, \textbf{b}, \textbf{c}) as a function of the rank used to approximate the signal evolution. The NRMSE for a rank 2 approximation is not displayed for $T_2$ since the values are too large for the displayed area. The $PD$- $T_1$- and $T_2$-to noise ratio in white matter are shown in (\textbf{d}, \textbf{e}, \textbf{f}) normalized by the input $SNR_{\text{in}}$.
}
\CaptRMSESNR                             

\subsection*{Figure \ref{fig:RMSE_spokes}}
\def\CaptRMSEspkoes{
	The normalized root mean square error (NRMSE) was calculated for the low rank ADMM algorithm when using 1 radial spoke (\textbf{a}) and 4 spokes per $TR$ (\textbf{b}), respectively. Note that the NRMSE of $T_1$ is bound by the discretization of the dictionary in (\textbf{b}). 
}
\CaptRMSEspkoes

\subsection*{Figure \ref{fig:Convergence}}
\def\CaptConvergence{
	The convergence of the low rank ADMM algorithm is visualized in terms of the normalized root mean square error (NRMSE) of the quantitative maps (\textbf{a},\textbf{b},\textbf{c}) and the residuum (Eq. \eqref{eq:MRF_LR_Problem}) (\textbf{d}). The reconstructions were performed with $R=5$. The color scale visualizes the number of ADMM iterations, where simulations with 10 iterations were highlighted by the color black and have been used for all other simulations in the present paper. The arrows indicate $\mu=1.26 \cdot 10^{-3}$, which was used for all simulations in this paper, if not stated otherwise.
}
\CaptConvergence

\subsection*{Figure \ref{fig:SVDRecon}}
\def\CaptSVDRecon{
	The series of images $\tilde{\mathbf{x}}$ reconstructed from noise free synthetic data are shown. The rows correspond to the singular values in decreasing order and their color coding corresponds to Fig.~\ref{fig:TimeSeries} (b, c). The columns show the different low rank (LR) reconstruction techniques discussed in this paper. Only the top half of the images are shown for better depiction. The right hand side of all images are scaled equally, while the left hand side is scaled equally only within the same row.
}
\CaptSVDRecon

\subsection*{Figure \ref{fig:Maps_nSVD_5}}
\def\CaptMapsF{
	The displayed excerpts of quantitative maps compare the discussed reconstruction methods. The simulations are based on a numerical phantom and the maps correspond to the singular value images shown in Fig.~\ref{fig:SVDRecon}. The maps were reconstructed from noise free data using $R=5$ singular values. 
}
\CaptMapsF

\subsection*{Figure \ref{fig:SNR}}
\def\CaptSNR{
	The mean value and the standard deviation (error bars) within fat, white matter (WM), gray matter (GM) and cerebrospinal fluid (CSF) are shown. The values are normalized by the original parameters used in the simulation. All low rank (LR) reconstructions were performed with $R=5$. Error bars are not displayed for those cases, where strong deviations from a Gaussian distribution make the calculation of the standard deviation questionable.
	}
\CaptSNR

\subsection*{Figure \ref{fig:Maps_nSVD_20}}
\def\CaptMapsT{
	The displayed excerpts of quantitative maps compare low rank inverse reconstructions with $R=20$. The simulations are based on a numerical phantom and were reconstructed from noise free data.
}
\CaptMapsT

\subsection*{Figure \ref{fig:SENSE}}
\def\CaptSENSE{
	The effect of parallel imaging is demonstrated at the example of a simulated data set with a matrix size of $384 \times 384$. A total of 841 radial spokes and in vivo coil sensitivity profiles of a 44 channel head coil (reduced to 12 virtual coils) were used. 
}
\CaptSENSE

\subsection*{Figure \ref{fig:InVivo}}
\def\CaptInVivo{
	The depicted parameter maps with a matrix size of $192 \times 192$  were reconstructed from 841 radial spokes, which were acquired in vivo within $4.1$~s. Coil sensitivity profiles of a 16 channel head coil (compressed to 8 virtual coil elements) were incorporated. The red rectangle at the top left indicates the region of interest in which mean and standard deviation of the relaxation times were calculated (Tab.~\ref{tab:in_vivo}). Note that the relaxation times are biased by slice profile imperfections and by magnetization transfer effects. 
}
\CaptInVivo

\subsection*{Figure \ref{fig:InVivo_32spokes}}
\def\CaptInVivoSpokes{
	The depicted parameter maps with a matrix size of $192 \times 192$  were reconstructed from $32\times 841$ radial spokes, which were acquired in vivo within $7.35$~minutes. Coil sensitivity profiles of a 16 channel head coil (compressed to 8 virtual coil elements) were incorporated. Note that the relaxation times are biased by $B_0$-drifts during the scan, by slice profile imperfections and by magnetization transfer effects. 
}
\CaptInVivoSpokes

\subsection*{Table \ref{tab:in_vivo}}
\def\CaptInVivoTable{
	Mean value and standard deviation of the relaxation times were calculated over a region of interest in the frontal white matter (Fig.~\ref{fig:InVivo}). Note that values calculated from 32 repetitions are strongly biased by $B_0$-drifts during the scan, while all values are affected by slice profile imperfections and magnetization transfer effects. 
}
\CaptInVivoTable

\def\colwidth{8.67cm}
\def\opfcolwidth{13.02cm}
\def\tcolwidth{17.56cm}

\newpage
\section{Figures}
\begin{figure}[h]
	\centering
	\if\submit1
	\includegraphics{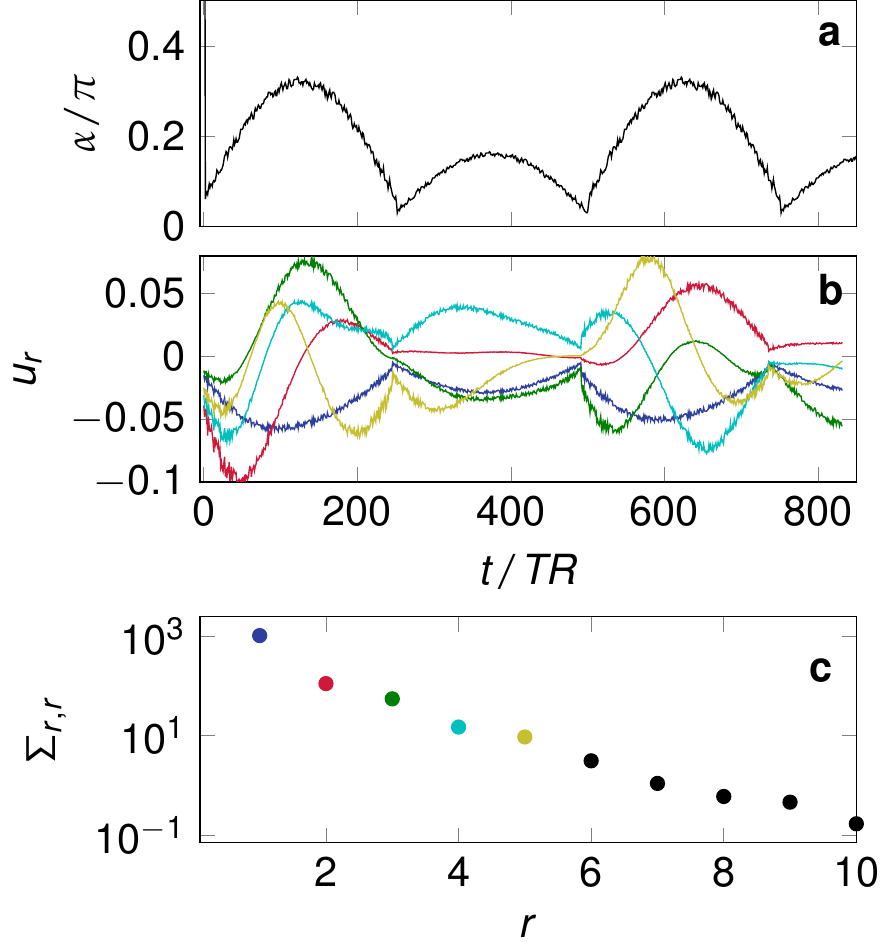}
	\else
	\begin{tikzpicture}
	\begin{axis}[
	width=\textwidth*0.5,
	height=\textheight*0.17,
	xmin=-5,
	xmax=850,
	ymin=0,
	ymax=0.5,
	/pgf/number format/.cd,	1000 sep={},
	ylabel={$\alpha/\pi$},
	name=alpha,
	xticklabel=\empty,
	]
	\addplot [color=black, solid]table[x=t, y=a]{Figures/alpha.dat};
	\node[right, inner sep=0mm] at (axis cs: 800,  .425) {\textbf{a}};
	\end{axis}
	
	\begin{axis}[
	width=\textwidth*0.5,
	height=\textheight*0.17,
	xmin=-5,
	xmax=850,
	ymin=-.1,
	ymax=.08,
	xlabel={$t/TR$},
	ylabel={$u_r$},
	at=(alpha.below south west),
	anchor=above north west,
	name=u,
	legend columns=2,
	yticklabel style={/pgf/number format/fixed},
	]
	
	\addplot [color=UKLblue, solid]             table[x=t, y=u1]{Figures/u.dat};
	\addplot [color=UKLred, solid]               table[x=t, y=u2]{Figures/u.dat};
	\addplot [color=green!50!black, solid]  table[x=t, y=u3]{Figures/u.dat};
	\addplot [color=turquois, solid]             table[x=t, y=u4]{Figures/u.dat};
	\addplot [color=yellow!75!black, solid]  table[x=t, y=u5]{Figures/u.dat};
	
	\node[right, inner sep=0mm] at (axis cs: 800,  .425/.5*.18-.1) {\textbf{b}};
	\end{axis}
	
	\begin{semilogyaxis}[
	width=\textwidth*0.5,
	height=\textheight*0.17,
	xmax=10,
	/pgf/number format/.cd,	1000 sep={},
	ylabel={$\Sigma_{r,r}$},
	xlabel={$r$},
	name=sv,
	at=(u.below south west),
	anchor=above north west,
	]
	\addplot [only marks, mark=*, scatter, scatter/classes={1={UKLblue}, 2={UKLred}, 3={green!50!black}, 4={turquois}, 5={yellow!75!black}, 6={black}}, scatter src=explicit] 
	table [x=n, y=sv,meta=color]{Figures/SV.dat};
	
	\node[right, inner sep=0mm] at (axis cs: 700/75.4,  200) {\textbf{c}};
	\end{semilogyaxis}
	\end{tikzpicture}%
	\fi
	\caption{\CaptTimeSeries}
	\label{fig:TimeSeries}
\end{figure}

\begin{figure}[h]
	\centering
	\if\submit1
	\includegraphics{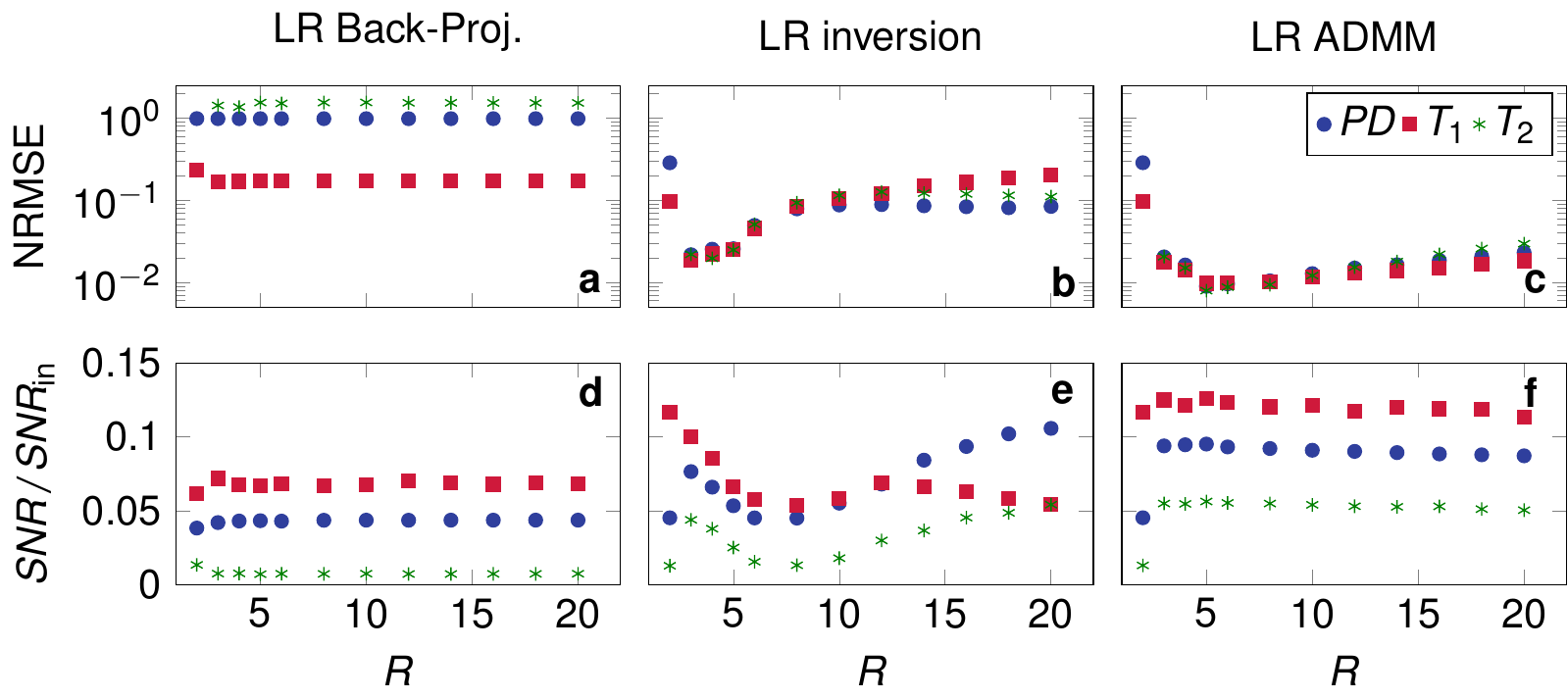}
	\else
	\begin{tikzpicture}
	\begin{semilogyaxis}[
	width=\textwidth*0.375,
	height=\textheight*0.17,
	xmin=1.01,
	xmax=22,
	ymin=.005,
	ymax=2.5,
	/pgf/number format/.cd,	1000 sep={},
	xticklabel=\empty,
	yticklabel style={/pgf/number format/fixed},
	ylabel={NRMSE},
	name=RMSE_SVDBP,
	title={LR Back-Proj.},
	]
	\addplot [only marks, mark=*, UKLblue] 	table [x=nsvd, y=PD_rmse]{Figures/RMSE_SVDBP.dat};
	\addplot [only marks, mark=square*, UKLred]    table [x=nsvd, y=T1_rmse]{Figures/RMSE_SVDBP.dat};
	\addplot [only marks, mark=asterisk, green!50!black] 	table [x=nsvd, y=T2_rmse]{Figures/RMSE_SVDBP.dat};
	
	\node[right, inner sep=0mm] at (axis cs: 20,  .01) {\textbf{a}};
	\end{semilogyaxis}
	
	\begin{semilogyaxis}[
	width=\textwidth*0.375,
	height=\textheight*0.17,
	xmin=1.01,
	xmax=22,
	ymin=.005,
	ymax=2.5,
	/pgf/number format/.cd,	1000 sep={},
	xticklabel=\empty,
	yticklabel=\empty,
	yticklabel style={/pgf/number format/fixed},
	name=RMSE_SVD,
	at=(RMSE_SVDBP.right of north east),
	anchor=left of north west,
	title={LR inversion},
	]
	\addplot [only marks, mark=*, UKLblue] 	table [x=nsvd, y=PD_rmse]{Figures/RMSE_SVD.dat};
	\addplot [only marks, mark=square*, UKLred]    table [x=nsvd, y=T1_rmse]{Figures/RMSE_SVD.dat};
	\addplot [only marks, mark=asterisk, green!50!black] 	table [x=nsvd, y=T2_rmse]{Figures/RMSE_SVD.dat};
	
	\node[right, inner sep=0mm] at (axis cs: 20,  .01) {\textbf{b}};
	\end{semilogyaxis}
	
	\begin{semilogyaxis}[
	width=\textwidth*0.375,
	height=\textheight*0.17,
	xmin=1.01,
	xmax=22,
	ymin=.005,
	ymax=2.5,
	/pgf/number format/.cd,	1000 sep={},
	xticklabel=\empty,
	yticklabel=\empty,
	name=RMSE_ADMM,
	at=(RMSE_SVD.right of north east),
	anchor=left of north west,
	title={LR ADMM},
	legend entries = {$PD$, $T_1$, $T_2$},
	legend pos = north east,
	legend columns=3, 
	]
	\addplot [only marks, mark=*, UKLblue] 	table [x=nsvd, y=PD_rmse]{Figures/RMSE_ADMM.dat};
	\addplot [only marks, mark=square*, UKLred]    table [x=nsvd, y=T1_rmse]{Figures/RMSE_ADMM.dat};
	\addplot [only marks, mark=asterisk, green!50!black] 	table [x=nsvd, y=T2_rmse]{Figures/RMSE_ADMM.dat};
	
	\node[right, inner sep=0mm] at (axis cs: 20,  .01) {\textbf{c}};
	
	\end{semilogyaxis}

	\begin{axis}[
	width=\textwidth*0.375,
	height=\textheight*0.17,
	xmin=1.01,
	xmax=22,
	ymin=0,
	ymax=.15,
	/pgf/number format/.cd,	1000 sep={},
	yticklabel style={/pgf/number format/fixed},
	xlabel={$R$},
	ylabel={$SNR/SNR_{\text{in}}$},
	name=SNR_SVDBP,
	at=(RMSE_SVDBP.below south east),
	anchor=above north east,
	]
	\addplot [only marks, mark=*, UKLblue] 	table [x=nsvd, y=PDNR]{Figures/SNR_SVDBP.dat};
	\addplot [only marks, mark=square*, UKLred]    table [x=nsvd, y=T1NR]{Figures/SNR_SVDBP.dat};
	\addplot [only marks, mark=asterisk, green!50!black] 	table [x=nsvd, y=T2NR]{Figures/SNR_SVDBP.dat};
	
	\node[right, inner sep=0mm] at (axis cs: 20,  .13) {\textbf{d}};
	\end{axis}
	
	\begin{axis}[
	width=\textwidth*0.375,
	height=\textheight*0.17,
	xmin=1.01,
	xmax=22,
	ymin=0,
	ymax=.15,
	/pgf/number format/.cd,	1000 sep={},
	yticklabel style={/pgf/number format/fixed},
	xlabel={$R$},
	yticklabel=\empty,
	name=SNR_SVD,
	at=(SNR_SVDBP.right of north east),
	anchor=left of north west,
	]
	\addplot [only marks, mark=*, UKLblue] 	table [x=nsvd, y=PDNR]{Figures/SNR_SVD.dat};
	\addplot [only marks, mark=square*, UKLred]    table [x=nsvd, y=T1NR]{Figures/SNR_SVD.dat};
	\addplot [only marks, mark=asterisk, green!50!black] 	table [x=nsvd, y=T2NR]{Figures/SNR_SVD.dat};
	
	\node[right, inner sep=0mm] at (axis cs: 20,  .13) {\textbf{e}};
	\end{axis}
	
	\begin{axis}[
	width=\textwidth*0.375,
	height=\textheight*0.17,
	xmin=1.01,
	xmax=21,
	ymin=0,
	ymax=.15,
	/pgf/number format/.cd,	1000 sep={},
	xlabel={$R$},
	yticklabel=\empty,
	xmax = 22,
	name=SNR_ADMM,
	at=(SNR_SVD.right of north east),
	anchor=left of north west,
	]
	\addplot [only marks, mark=*, UKLblue] 	table [x=nsvd, y=PDNR]{Figures/SNR_ADMM.dat};
	\addplot [only marks, mark=square*, UKLred]    table [x=nsvd, y=T1NR]{Figures/SNR_ADMM.dat};
	\addplot [only marks, mark=asterisk, green!50!black] 	table [x=nsvd, y=T2NR]{Figures/SNR_ADMM.dat};
	
	\node[right, inner sep=0mm] at (axis cs: 20,  .13) {\textbf{f}};
	
	\end{axis}
	\end{tikzpicture}%
	\fi
	\caption{\CaptRMSESNR}
	\label{fig:RMSE_SNR}
\end{figure}

\begin{figure}[h]
	\centering
	\if\submit1
	\includegraphics{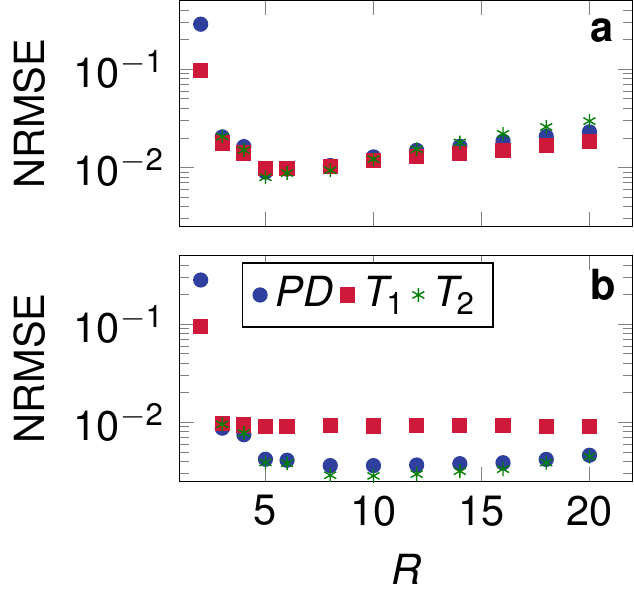}
	\else
	\begin{tikzpicture}
	\begin{semilogyaxis}[
	width=\textwidth*0.375,
	height=\textheight*0.17,
	xmin=1.01,
	xmax=22,
	ymin=.0025,
	ymax=0.5,
	/pgf/number format/.cd,	1000 sep={},
	xticklabel=\empty,
	yticklabel style={/pgf/number format/fixed},
	ylabel={NRMSE},
	name=RMSE1,
	]
	\addplot [only marks, mark=*, UKLblue] 	table [x=nsvd, y=PD_rmse]{Figures/RMSE_ADMM.dat};
	\addplot [only marks, mark=square*, UKLred]    table [x=nsvd, y=T1_rmse]{Figures/RMSE_ADMM.dat};
	\addplot [only marks, mark=asterisk, green!50!black] 	table [x=nsvd, y=T2_rmse]{Figures/RMSE_ADMM.dat};
	
	\node[right, inner sep=0mm] at (axis cs: 20,  .25) {\textbf{a}};
	\end{semilogyaxis}
	
	\begin{semilogyaxis}[
	width=\textwidth*0.375,
	height=\textheight*0.17,
	xmin=1.01,
	xmax=22,
	ymin=.0025,
	ymax=0.5,
	/pgf/number format/.cd,	1000 sep={},
	xlabel={$R$},
	ylabel={NRMSE},
	name=RMSE2,
	at=(RMSE1.below south east),
	anchor=above north east,
	legend entries = {$PD$, $T_1$, $T_2$},
	legend pos = north west,
	legend columns=3, 
	legend style={xshift=0.5cm},
	]
	\addplot [only marks, mark=*, UKLblue] 	table [x=nsvd, y=PD_rmse]{Figures/RMSE_ADMM_4_repetitions.dat};
	\addplot [only marks, mark=square*, UKLred]    table [x=nsvd, y=T1_rmse]{Figures/RMSE_ADMM_4_repetitions.dat};
	\addplot [only marks, mark=asterisk, green!50!black] 	table [x=nsvd, y=T2_rmse]{Figures/RMSE_ADMM_4_repetitions.dat};
	
	\node[right, inner sep=0mm] at (axis cs: 20,  .25) {\textbf{b}};
	
	\end{semilogyaxis}
	
%
%
	\end{tikzpicture}%
	\fi
	\caption{\CaptRMSEspkoes}
	\label{fig:RMSE_spokes}
\end{figure}

\begin{figure}[h]
	\centering
	\if\submit1
	\includegraphics{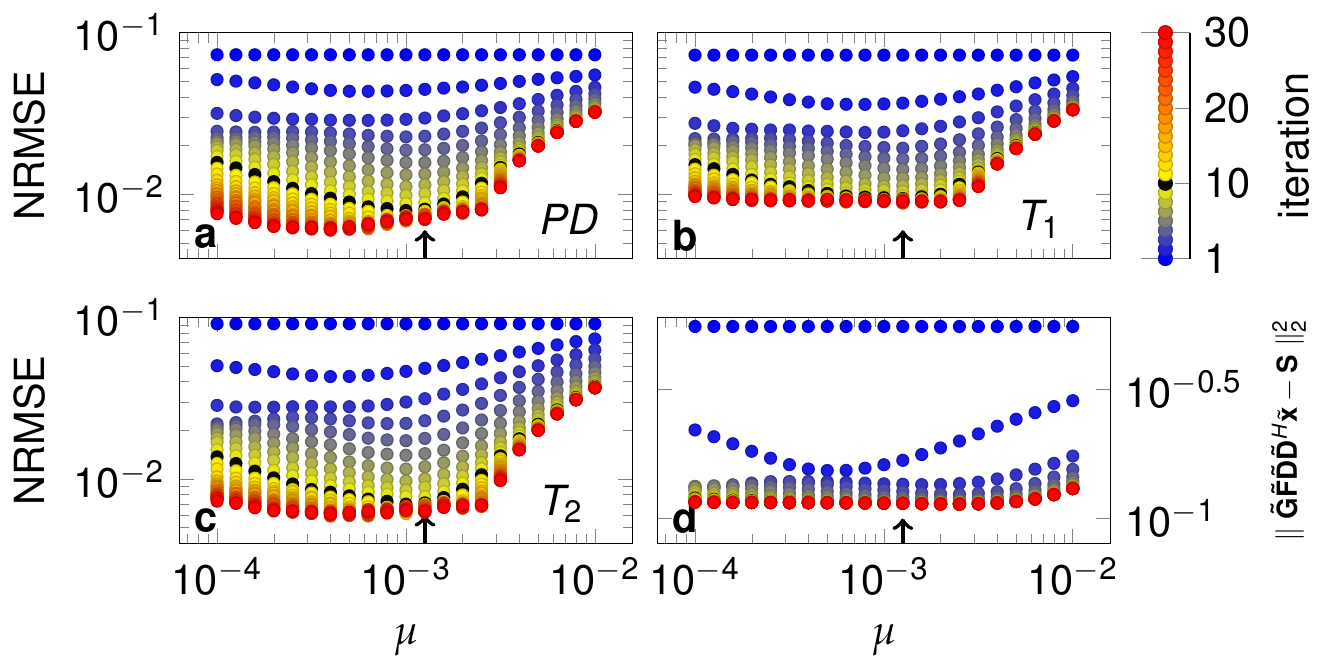}
	\else
	\begin{tikzpicture}
	\begin{loglogaxis}[
	width=\textwidth*0.375,
	height=\textheight*0.17,
	ymin=.004,
	ymax=.1,
	/pgf/number format/.cd,	1000 sep={},
	xticklabel=\empty,
	ylabel={NRMSE},
	name={PD_nrmse},
	colormap name = hotblack,
	]
	\addplot [only marks, mark=*, scatter, point meta=explicit, every mark/.append style={scale=.85},] 	table [x=mu, y=PD_rmse, meta=iteration]{Figures/convergence_mu.dat};
	
	\node[right, inner sep=0mm] at (axis cs: .000075,  0.0055) {\textbf{a}};
	\node[right, inner sep=0mm] at (axis cs: .005,  0.007) {$PD$};
	\draw[->, very thick](axis cs: 0.00126, 0.004) -- (axis cs: 0.00126, 0.006);
	\end{loglogaxis}
	
	\begin{loglogaxis}[
	width=\textwidth*0.375,
	height=\textheight*0.17,
	ymin=.004,
	ymax=.1,
	yticklabel=\empty,
	xticklabel=\empty,
	name=T1_nrmse,
	at=(PD_nrmse.north east),
	anchor=north west,
	xshift = .25cm,
	colorbar sampled line,
	colorbar style={ylabel=iteration, ytick={0,10,20,30}, yticklabels={$1$, $10$, $20$, $30$}},
	colormap name = hotblack,
	]
	\addplot [only marks, mark=*, scatter, point meta=explicit, every mark/.append style={scale=.85},] 	table [x=mu, y=T1_rmse, meta=iteration]{Figures/convergence_mu.dat};
	
	\node[right, inner sep=0mm] at (axis cs: .000075,  0.0055) {\textbf{b}};
	\node[right, inner sep=0mm] at (axis cs: .005,  0.007) {$T_1$};
	\draw[->, very thick](axis cs: 0.00126, 0.004) -- (axis cs: 0.00126, 0.006);
	\end{loglogaxis}
	
	\begin{loglogaxis}[
	width=\textwidth*0.375,
	height=\textheight*0.17,
	ymin=.004,
	ymax=.1,
	xlabel={$\mu$},
	ylabel={NRMSE},
	name=T2_nrmse,
	at=(PD_nrmse.south east),
	anchor=north east,
	yshift = -.6cm,
	colormap name = hotblack,
	]
	\addplot [only marks, mark=*, scatter, point meta=explicit, every mark/.append style={scale=.85},] 	table [x=mu, y=T2_rmse, meta=iteration]{Figures/convergence_mu.dat};
	
	\node[right, inner sep=0mm] at (axis cs: .000075,  0.0055) {\textbf{c}};
	\node[right, inner sep=0mm] at (axis cs: .005,  0.007) {$T_2$};
	\draw[->, very thick](axis cs: 0.00126, 0.004) -- (axis cs: 0.00126, 0.006);
	\end{loglogaxis}
	
	\begin{loglogaxis}[
	width=\textwidth*0.375,
	height=\textheight*0.17,
	ymin= 0.08,
	ymax=0.6,
	xlabel={$\mu$},
	ylabel={\scriptsize$\parallel \mathbf{\tilde{G}} \mathbf{\tilde{F}} \mathbf{\tilde{D}} \mathbf{\tilde{D}}^H \tilde{\mathbf{x}} - \mathbf{S} \parallel_2^2$},
	yticklabel pos=right,
	at=(PD_nrmse.south east),
	anchor=north west,
	xshift = .25cm,
	yshift = -.6cm,
	colormap name = hotblack,
	]
	\addplot [only marks, mark=*, scatter, point meta=explicit, every mark/.append style={scale=.85},] 	table [x=mu, y=residuum, meta=iteration]{Figures/convergence_mu.dat};
	\node[right, inner sep=0mm] at (axis cs: .000075,  0.1) {\textbf{d}};
	\draw[->, very thick](axis cs: 0.00126, 0.08) -- (axis cs: 0.00126, 0.1);
	\end{loglogaxis}
	\end{tikzpicture}%
	\fi
	\caption{\CaptConvergence}
	\label{fig:Convergence}
\end{figure}

\begin{figure}[hp]
	\centering
	\if\submit1
	\includegraphics{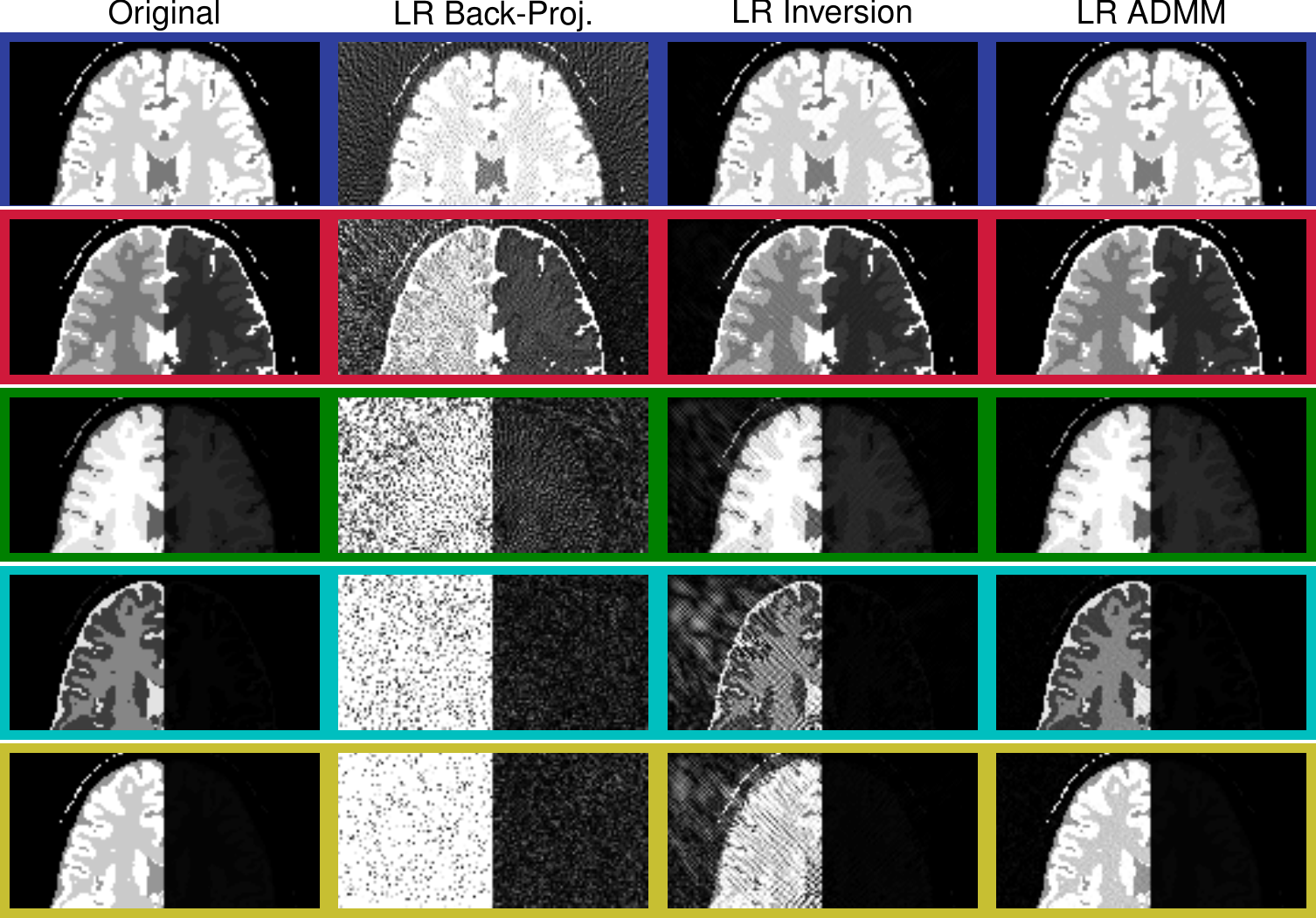}
	\else
	
	\begin{tikzpicture}[scale=.9]
	\begin{axis}[%
	width=17cm,
	height=1cm,
	scale only axis,
	xmin=-0.125,
	xmax=16.875,
	ymin=0,
	ymax=1,
	hide axis,
	name=title]
	]	
	\node[above, inner sep=0mm] at (axis cs: 2, 0) {Original};
	\node[above, inner sep=0mm] at (axis cs: 6.25,  0) {LR Back-Proj.};
	\node[above, inner sep=0mm] at (axis cs:10.5,  .1) {LR Inversion};
	\node[above, inner sep=0mm] at (axis cs:14.75,  .1) {LR ADMM};
	\end{axis}
	
	\begin{axis}[%
	width=17cm,
	height=2.25cm,
	scale only axis,
	xmin=-0.125,
	xmax=16.875,
	ymin=-.0125,
	ymax=2.125,
	hide axis,
	name=SVD1,
	at=(title.south east),
	anchor=north east,
	axis background/.style={fill=UKLblue}]
	]
	\addplot graphics [xmin=0,xmax=4,ymin=0,ymax=2] {Figures/SVD_org_1.png};
	\addplot graphics [xmin=4.25,xmax=8.25,ymin=0,ymax=2] {Figures/SVD_BP_1.png};
	\addplot graphics [xmin=8.5,xmax=12.5,ymin=0,ymax=2] {Figures/SVD_1.png};
	\addplot graphics [xmin=12.75,xmax=16.75,ymin=0,ymax=2] {Figures/SVD_ADMM_1.png};
	\end{axis}
	
	\begin{axis}[%
	width=17cm,
	height=2.25cm,
	scale only axis,
	xmin=-0.125,
	xmax=16.875,
	ymin=-.125,
	ymax=2.125,
	hide axis,
	name=SVD2,
	at=(SVD1.south east),
	anchor=north east,
	yshift = -.05cm,
	axis background/.style={fill=UKLred}]
	]
	\addplot graphics [xmin=0,xmax=4,ymin=0,ymax=2] {Figures/SVD_org_2.png};
	\addplot graphics [xmin=4.25,xmax=8.25,ymin=0,ymax=2] {Figures/SVD_BP_2.png};
	\addplot graphics [xmin=8.5,xmax=12.5,ymin=0,ymax=2] {Figures/SVD_2.png};
	\addplot graphics [xmin=12.75,xmax=16.75,ymin=0,ymax=2] {Figures/SVD_ADMM_2.png};
	\end{axis}
	
	\begin{axis}[%
	width=17cm,
	height=2.25cm,
	scale only axis,
	xmin=-0.125,
	xmax=16.875,
	ymin=-.125,
	ymax=2.125,
	hide axis,
	name=SVD3,
	at=(SVD2.south east),
	anchor=north east,
	yshift = -.05cm,
	axis background/.style={fill=green!50!black}]
	]
	\addplot graphics [xmin=0,xmax=4,ymin=0,ymax=2] {Figures/SVD_org_3.png};
	\addplot graphics [xmin=4.25,xmax=8.25,ymin=0,ymax=2] {Figures/SVD_BP_3.png};
	\addplot graphics [xmin=8.5,xmax=12.5,ymin=0,ymax=2] {Figures/SVD_3.png};
	\addplot graphics [xmin=12.75,xmax=16.75,ymin=0,ymax=2] {Figures/SVD_ADMM_3.png};
	\end{axis}
	
	\begin{axis}[%
	width=17cm,
	height=2.25cm,
	scale only axis,
	xmin=-0.125,
	xmax=16.875,
	ymin=-.125,
	ymax=2.125,
	hide axis,
	name=SVD4,
	at=(SVD3.south east),
	anchor=north east,
	yshift = -.05cm,
	axis background/.style={fill=turquois}]
	]
	\addplot graphics [xmin=0,xmax=4,ymin=0,ymax=2] {Figures/SVD_org_4.png};
	\addplot graphics [xmin=4.25,xmax=8.25,ymin=0,ymax=2] {Figures/SVD_BP_4.png};
	\addplot graphics [xmin=8.5,xmax=12.5,ymin=0,ymax=2] {Figures/SVD_4.png};
	\addplot graphics [xmin=12.75,xmax=16.75,ymin=0,ymax=2] {Figures/SVD_ADMM_4.png};
	\end{axis}
	
	\begin{axis}[%
	width=17cm,
	height=2.25cm,
	scale only axis,
	xmin=-0.125,
	xmax=16.875,
	ymin=-.125,
	ymax=2.125,
	hide axis,
	name=SVD5,
	at=(SVD4.south east),
	anchor=north east,
	yshift = -.05cm,
	axis background/.style={fill=yellow!75!black}]
	]
	\addplot graphics [xmin=0,xmax=4,ymin=0,ymax=2] {Figures/SVD_org_5.png};
	\addplot graphics [xmin=4.25,xmax=8.25,ymin=0,ymax=2] {Figures/SVD_BP_5.png};
	\addplot graphics [xmin=8.5,xmax=12.5,ymin=0,ymax=2] {Figures/SVD_5.png};
	\addplot graphics [xmin=12.75,xmax=16.75,ymin=0,ymax=2] {Figures/SVD_ADMM_5.png};
	\end{axis}
	\end{tikzpicture}
	\fi
	\caption{\CaptSVDRecon}
	\label{fig:SVDRecon}
\end{figure}

\begin{figure}[hp]
	\centering
	\if\submit1
	\includegraphics{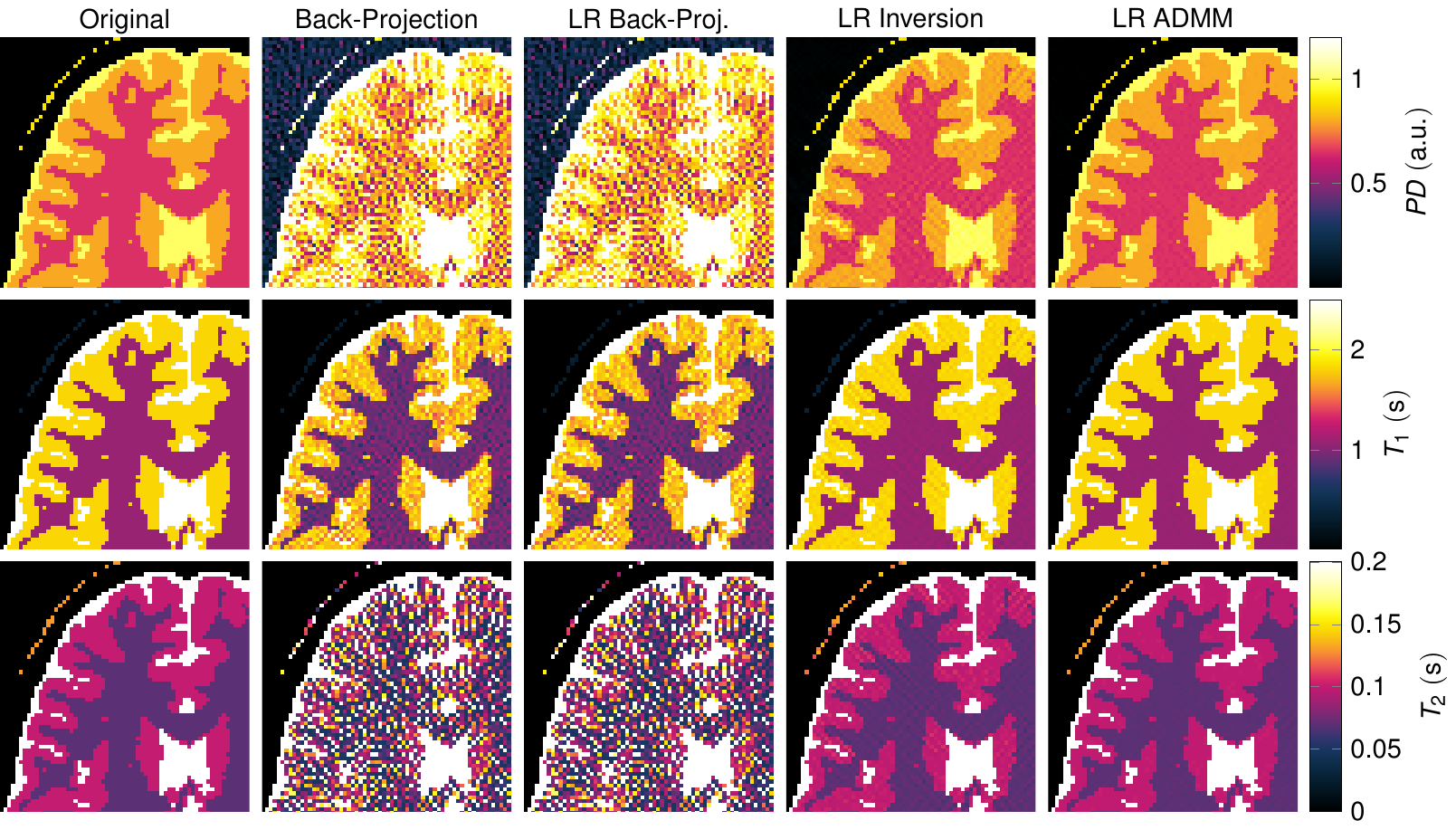}
	\else
	
	\begin{tikzpicture}[scale=.7]
	\begin{axis}[%
	width=4cm,
	height=4cm,
	axis on top,
	scale only axis,
	xmin=0,
	xmax=1,
	y dir=reverse,
	ymin=0,
	ymax=1,
	hide axis,
	name=PDOrg,
	title={Original},
	title style={yshift = -.3cm},
	colormap name = morgenstemning,
	point meta min=0,
	point meta max=1,	
	]
	\addplot graphics [xmin=-0.25,xmax=1.75,ymin=0,ymax=2] {Figures/PD_org.png};
	\end{axis}
	
	\begin{axis}[%
	width=4cm,
	height=4cm,
	axis on top,
	scale only axis,
	xmin=0,
	xmax=1,
	y dir=reverse,
	ymin=0,
	ymax=1,
	hide axis,
	name=PDMRF,
	title={Back-Projection},
	title style={yshift = -.3cm},
	colormap name = morgenstemning,
	point meta min=0.01,
	point meta max=1.2,
	at=(PDOrg.right of north east),
	anchor=left of north west,
	xshift=.08in,
	]
	\addplot graphics [xmin=-0.25,xmax=1.75,ymin=0,ymax=2] {Figures/PD_backprojection.png};
	\end{axis}
	
	\begin{axis}[%
	width=4cm,
	height=4cm,
	axis on top,
	scale only axis,
	xmin=0,
	xmax=1,
	y dir=reverse,
	ymin=0,
	ymax=1,
	hide axis,
	name=PDSVDBP,
	title={LR Back-Proj.},
	title style={yshift = -.3cm},
	colormap name = morgenstemning,
	point meta min=0.01,
	point meta max=1.2,
	at=(PDMRF.right of north east),
	anchor=left of north west,
	xshift=.08in,
	]
	\addplot graphics [xmin=-0.25,xmax=1.75,ymin=0,ymax=2] {Figures/PD_nSVD_5_backprojection.png};
	\end{axis}
	
	\begin{axis}[%
	width=4cm,
	height=4cm,
	axis on top,
	scale only axis,
	xmin=0,
	xmax=1,
	y dir=reverse,
	ymin=0,
	ymax=1,
	hide axis,
	name=PDSVD,
	title={LR Inversion},
	title style={yshift = -.2cm},
	colormap name = morgenstemning,
	point meta min=0.01,
	point meta max=1.2,
	at=(PDSVDBP.right of north east),
	anchor=left of north west,
	xshift=.08in,
	]
	\addplot graphics [xmin=-0.25,xmax=1.75,ymin=0,ymax=2] {Figures/PD_nSVD_5.png};
	\end{axis}
	
	\begin{axis}[%
	width=4cm,
	height=4cm,
	axis on top,
	scale only axis,
	xmin=0,
	xmax=1,
	y dir=reverse,
	ymin=0,
	ymax=1,
	hide axis,
	name=PDADMM,
	title={LR ADMM},
	title style={yshift = -.2cm},
	colormap name = morgenstemning,
	colorbar,
	point meta min=0.001,
	point meta max=1.2,
	at=(PDSVD.right of north east),
	anchor=left of north west,
	xshift=.08in,
	colorbar style={ylabel=$PD~(\text{a.u.})$, width=0.5cm, xshift=-0.1cm},
	]
	\addplot graphics [xmin=-0.25,xmax=1.75,ymin=0,ymax=2] {Figures/PD_ADMM_nSVD_5.png};
	\end{axis}
	
	\begin{axis}[%
	width=4cm,
	height=4cm,
	axis on top,
	scale only axis,
	xmin=0,
	xmax=1,
	y dir=reverse,
	ymin=0,
	ymax=1,
	hide axis,
	name=T1Org,
	colormap name = morgenstemning,
	at=(PDOrg.south east),
	anchor=north east,
	yshift = -.08in,
	]
	\addplot graphics [xmin=-0.25,xmax=1.75,ymin=0,ymax=2] {Figures/T1_org.png};
	\end{axis}
	
	\begin{axis}[%
	width=4cm,
	height=4cm,
	axis on top,
	scale only axis,
	xmin=0,
	xmax=1,
	y dir=reverse,
	ymin=0,
	ymax=1,
	hide axis,
	name=T1MRF,
	colormap name = morgenstemning,
	point meta min=0.01,
	point meta max=1.5,
	at=(T1Org.right of north east),
	anchor=left of north west,
	xshift=.08in,
	]
	\addplot graphics [xmin=-0.25,xmax=1.75,ymin=0,ymax=2] {Figures/T1_backprojection.png};
	
	\end{axis}
	
	\begin{axis}[%
	width=4cm,
	height=4cm,
	axis on top,
	scale only axis,
	xmin=0,
	xmax=1,
	y dir=reverse,
	ymin=0,
	ymax=1,
	hide axis,
	name=T1SVDBP,
	colormap name = morgenstemning,
	point meta min=0.01,
	point meta max=2.5,
	at=(T1MRF.right of north east),
	anchor=left of north west,
	xshift=.08in,
	]
	\addplot graphics [xmin=-0.25,xmax=1.75,ymin=0,ymax=2] {Figures/T1_nSVD_5_backprojection.png};
	\end{axis}
	
	\begin{axis}[%
	width=4cm,
	height=4cm,
	axis on top,
	scale only axis,
	xmin=0,
	xmax=1,
	y dir=reverse,
	ymin=0,
	ymax=1,
	hide axis,
	name=T1SVD,
	colormap name = morgenstemning,
	point meta min=0.01,
	point meta max=1.5,
	at=(T1SVDBP.right of north east),
	anchor=left of north west,
	xshift=.08in,
	]
	\addplot graphics [xmin=-0.25,xmax=1.75,ymin=0,ymax=2] {Figures/T1_nSVD_5.png};
	\end{axis}
	
	\begin{axis}[%
	width=4cm,
	height=4cm,
	axis on top,
	scale only axis,
	xmin=0,
	xmax=1,
	y dir=reverse,
	ymin=0,
	ymax=1,
	hide axis,
	name=T1ADMM,
	colormap name = morgenstemning,
	colorbar,
	point meta min=0.01,
	point meta max=2.5,
	at=(T1SVD.right of north east),
	anchor=left of north west,
	xshift=.08in,
	colorbar style={ylabel=$T_1~(\text{s})$, width=0.5cm, xshift=-0.1cm},
	]
	\addplot graphics [xmin=-0.25,xmax=1.75,ymin=0,ymax=2] {Figures/T1_ADMM_nSVD_5.png};
	\end{axis}
	
	\begin{axis}[%
	width=4cm,
	height=4cm,
	axis on top,
	scale only axis,
	xmin=0,
	xmax=1,
	y dir=reverse,
	ymin=0,
	ymax=1,
	hide axis,
	name=T2Org,
	colormap name = morgenstemning,
	at=(T1Org.south east),
	anchor=north east,
	yshift = -.08in,
	]
	\addplot graphics [xmin=-0.25,xmax=1.75,ymin=0,ymax=2] {Figures/T2_org.png};
	\end{axis}
	
	\begin{axis}[%
	width=4cm,
	height=4cm,
	axis on top,
	scale only axis,
	xmin=0,
	xmax=1,
	y dir=reverse,
	ymin=0,
	ymax=1,
	hide axis,
	name=T2MRF,
	colormap name = morgenstemning,
	point meta min=0,
	point meta max=0.3,
	at=(T2Org.right of north east),
	anchor=left of north west,
	xshift=.08in,
	]
	\addplot graphics [xmin=-0.25,xmax=1.75,ymin=0,ymax=2] {Figures/T2_backprojection.png};
	\end{axis}
	
	\begin{axis}[%
	width=4cm,
	height=4cm,
	axis on top,
	scale only axis,
	xmin=0,
	xmax=1,
	y dir=reverse,
	ymin=0,
	ymax=1,
	hide axis,
	name=T2SVDBP,
	colormap name = morgenstemning,
	point meta min=0,
	point meta max=0.3,
	at=(T2MRF.right of north east),
	anchor=left of north west,
	xshift=.08in,
	]
	\addplot graphics [xmin=-0.25,xmax=1.75,ymin=0,ymax=2] {Figures/T2_nSVD_5_backprojection.png};
	\end{axis}
	
	\begin{axis}[%
	width=4cm,
	height=4cm,
	axis on top,
	scale only axis,
	xmin=0,
	xmax=1,
	y dir=reverse,
	ymin=0,
	ymax=1,
	hide axis,
	name=T2SVD,
	colormap name = morgenstemning,
	point meta min=0,
	point meta max=0.3,
	at=(T2SVDBP.right of north east),
	anchor=left of north west,
	xshift=.08in,
	]
	\addplot graphics [xmin=-0.25,xmax=1.75,ymin=0,ymax=2] {Figures/T2_nSVD_5.png};
	\end{axis}
	
	\begin{axis}[%
	width=4cm,
	height=4cm,
	axis on top,
	scale only axis,
	xmin=0,
	xmax=1,
	y dir=reverse,
	ymin=0,
	ymax=1,
	hide axis,
	name=T2ADMM,
	colormap name = morgenstemning,
	colorbar,
	point meta min=0,
	point meta max=0.2,
	at=(T2SVD.right of north east),
	anchor=left of north west,
	xshift=.08in,
	colorbar style={ylabel=$T_2~(\text{s})$, width=0.5cm, xshift=-0.1cm, yticklabel style={/pgf/number format/fixed}},
	]
	\addplot graphics [xmin=-0.25,xmax=1.75,ymin=0,ymax=2] {Figures/T2_ADMM_nSVD_5.png};
	\end{axis}
	\end{tikzpicture}
	\fi
	\caption{\CaptMapsF}
	\label{fig:Maps_nSVD_5}
\end{figure}

\begin{figure}[hp]
	\centering
	\if\submit1
	\includegraphics{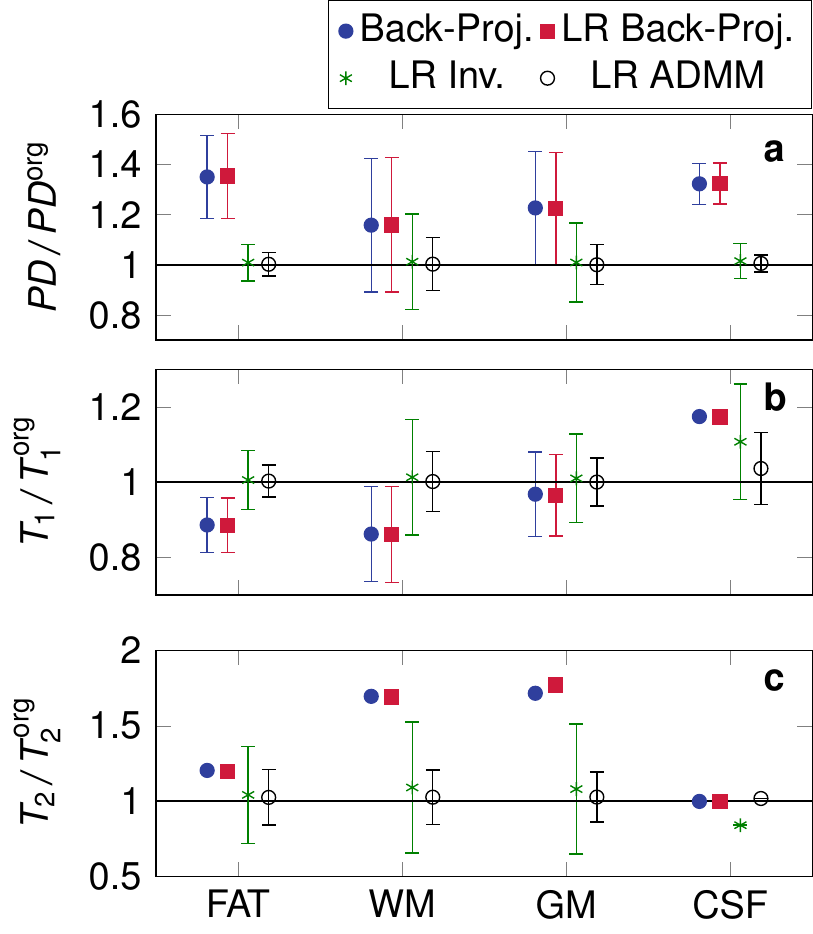}
	\else
	\begin{tikzpicture}[font=\small]
	\begin{axis}[
	width=\textwidth*0.5,
	height=\textheight*0.17,
	/pgf/number format/.cd,	1000 sep={},
	xticklabel=\empty,
	yticklabel style={/pgf/number format/fixed},
	xmin = 0.5,
	xmax = 4.5,
	ymin = 0.7,
	ymax = 1.6,
	ylabel={$PD/PD^{\text{org}}$},
	name=PD,
	legend entries = {Back-Proj., LR Back-Proj., LR Inv., LR ADMM},
	legend style={at={(axis cs:4.5,1.62)}, anchor=south east},
	legend columns=2, 
	]
	\addplot [only marks, mark=*, UKLblue, error bars/.cd,y dir=both, y explicit] 	table [x=x, y=PDmean, y error=PDerror]{Figures/noise_nsvd_5_MRF.dat};
	\addplot [only marks, mark=square*, UKLred,  error bars/.cd,y dir=both, y explicit] 	table [x=x, y=PDmean, y error=PDerror]{Figures/noise_nsvd_5_SVDBP.dat};
	\addplot [only marks, mark=asterisk, green!50!black, error bars/.cd,y dir=both, y explicit] 	table [x=x, y=PDmean, y error=PDerror]{Figures/noise_nsvd_5_SVD.dat};
	\addplot [only marks, mark=o, black, error bars/.cd,y dir=both, y explicit] 	table [x=x, y=PDmean, y error=PDerror]{Figures/noise_nsvd_5_ADMM.dat};
	
	\addplot[black, no marks] coordinates {(0,1) (5,1)};
	\node[right, inner sep=0mm] at (axis cs: 4.2,  1.45) {\textbf{a}};
	\end{axis}
	
	\begin{axis}[
	width=\textwidth*0.5,
	height=\textheight*0.17,
	/pgf/number format/.cd,	1000 sep={},
	xticklabel=\empty,
	yticklabel style={/pgf/number format/fixed},
	xmin = 0.5,
	xmax = 4.5,
	ymin = 0.7,
	ymax = 1.3,
	ylabel={$T_1/T_1^{\text{org}}$},
	name=T1,
	at=(PD.below south east),
	anchor=above north east,	
	]
	\addplot [only marks, mark=*, UKLblue, error bars/.cd,y dir=both, y explicit] 	table [x=x, y=T1mean, y error=T1error]{Figures/noise_nsvd_5_MRF.dat};
	\addplot [only marks, mark=square*, UKLred,  error bars/.cd,y dir=both, y explicit] 	table [x=x, y=T1mean, y error=T1error]{Figures/noise_nsvd_5_SVDBP.dat};
	\addplot [only marks, mark=asterisk, green!50!black, error bars/.cd,y dir=both, y explicit] 	table [x=x, y=T1mean, y error=T1error]{Figures/noise_nsvd_5_SVD.dat};
	\addplot [only marks, mark=o, black, error bars/.cd,y dir=both, y explicit] 	table [x=x, y=T1mean, y error=T1error]{Figures/noise_nsvd_5_ADMM.dat};
	
	\addplot[black, no marks] coordinates {(0,1) (5,1)};
	\node[right, inner sep=0mm] at (axis cs: 4.2,  1.225) {\textbf{b}};
	
	\end{axis}
	
	\begin{axis}[
	width=\textwidth*0.5,
	height=\textheight*0.17,
	/pgf/number format/.cd,	1000 sep={},
	xtick={1,2,3,4},
	xticklabels={FAT, WM, GM, CSF},
	yticklabel style={/pgf/number format/fixed},
	xmin = 0.5,
	xmax = 4.5,
	ymin = .5,
	ymax = 2,
	ylabel={$T_2/T_2^{\text{org}}$},
	name=T2,
	at=(T1.below south east),
	anchor=above north east,	
	]
	\addplot [only marks, mark=*, UKLblue, error bars/.cd,y dir=both, y explicit] 	table [x=x, y=T2mean, y error=T2error]{Figures/noise_nsvd_5_MRF.dat};
	\addplot [only marks, mark=square*, UKLred,  error bars/.cd,y dir=both, y explicit] 	table [x=x, y=T2mean, y error=T2error]{Figures/noise_nsvd_5_SVDBP.dat};
	\addplot [only marks, mark=asterisk, green!50!black, error bars/.cd,y dir=both, y explicit] 	table [x=x, y=T2mean, y error=T2error]{Figures/noise_nsvd_5_SVD.dat};
	\addplot [only marks, mark=o, black, error bars/.cd,y dir=both, y explicit] 	table [x=x, y=T2mean, y error=T2error]{Figures/noise_nsvd_5_ADMM.dat};
	
	\addplot[black, no marks] coordinates {(0,1) (5,1)};
	
	\node[right, inner sep=0mm] at (axis cs: 4.2,  1.8) {\textbf{c}};
	
	\end{axis}
	\end{tikzpicture}%
	\fi
	\caption{\CaptSNR}
	\label{fig:SNR}
\end{figure}

\begin{figure}[hp]
	\centering
	\if\submit1
	\includegraphics{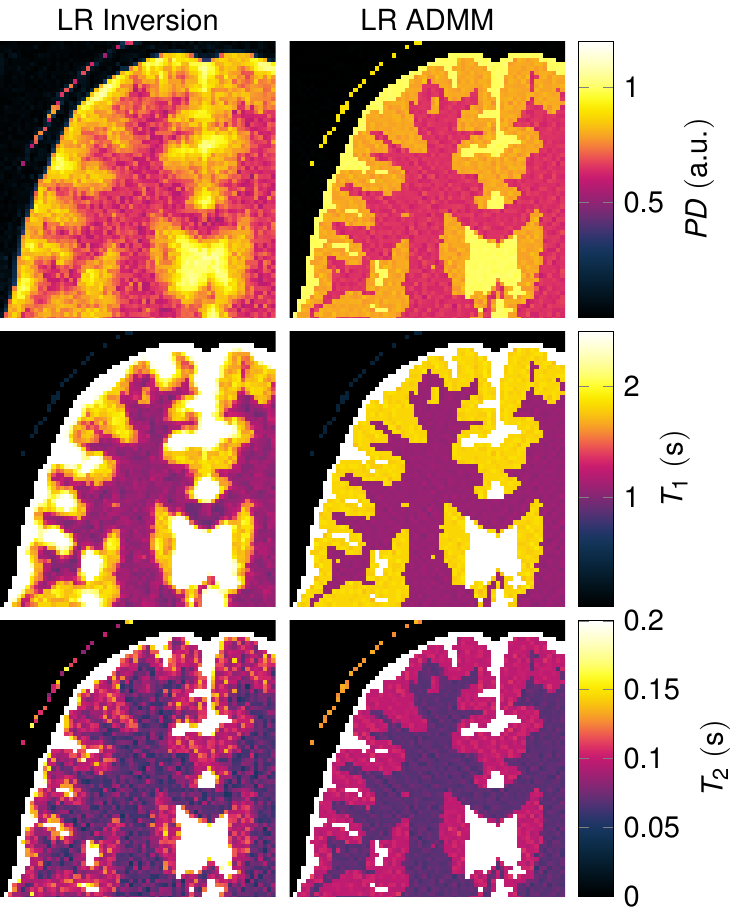}
	\else
	\begin{tikzpicture}[scale=.7]
	\begin{axis}[%
	width=4cm,
	height=4cm,
	axis on top,
	scale only axis,
	xmin=0,
	xmax=1,
	y dir=reverse,
	ymin=0,
	ymax=1,
	hide axis,
	name=PDSVD,
	title={LR Inversion},
	title style={yshift = -.2cm},
	colormap name = morgenstemning,
	point meta min=0.01,
	point meta max=1.2,
	xshift=.08in,
	]
	\addplot graphics [xmin=-0.25,xmax=1.75,ymin=0,ymax=2] {Figures/PD_nSVD_20.png};
	\end{axis}
	
	\begin{axis}[%
	width=4cm,
	height=4cm,
	axis on top,
	scale only axis,
	xmin=0,
	xmax=1,
	y dir=reverse,
	ymin=0,
	ymax=1,
	hide axis,
	name=PDADMM,
	title={LR ADMM},
	title style={yshift = -.2cm},
	colormap name = morgenstemning,
	colorbar,
	point meta min=0.001,
	point meta max=1.2,
	at=(PDSVD.right of north east),
	anchor=left of north west,
	xshift=.08in,
	colorbar style={ylabel=$PD~(\text{a.u.})$, width=0.5cm, xshift=-0.1cm},
	]
	\addplot graphics [xmin=-0.25,xmax=1.75,ymin=0,ymax=2] {Figures/PD_ADMM_nSVD_20.png};
	\end{axis}
	
	\begin{axis}[%
	width=4cm,
	height=4cm,
	axis on top,
	scale only axis,
	xmin=0,
	xmax=1,
	y dir=reverse,
	ymin=0,
	ymax=1,
	hide axis,
	name=T1SVD,
	colormap name = morgenstemning,
	point meta min=0.01,
	point meta max=1.5,
	at=(PDSVD.below south west),
	anchor=left of north west,
	yshift=-.08in,
	]
	\addplot graphics [xmin=-0.25,xmax=1.75,ymin=0,ymax=2] {Figures/T1_nSVD_20.png};
	\end{axis}
	
	\begin{axis}[%
	width=4cm,
	height=4cm,
	axis on top,
	scale only axis,
	xmin=0,
	xmax=1,
	y dir=reverse,
	ymin=0,
	ymax=1,
	hide axis,
	name=T1ADMM,
	colormap name = morgenstemning,
	colorbar,
	point meta min=0.01,
	point meta max=2.5,
	at=(T1SVD.right of north east),
	anchor=left of north west,
	xshift=.08in,
	colorbar style={ylabel=$T_1~(\text{s})$, width=0.5cm, xshift=-0.1cm},
	]
	\addplot graphics [xmin=-0.25,xmax=1.75,ymin=0,ymax=2] {Figures/T1_ADMM_nSVD_20.png};
	\end{axis}
	
	\begin{axis}[%
	width=4cm,
	height=4cm,
	axis on top,
	scale only axis,
	xmin=0,
	xmax=1,
	y dir=reverse,
	ymin=0,
	ymax=1,
	hide axis,
	name=T2SVD,
	colormap name = morgenstemning,
	point meta min=0,
	point meta max=0.3,
	at=(T1SVD.below south west),
	anchor=left of north west,
	yshift=-.08in,
	]
	\addplot graphics [xmin=-0.25,xmax=1.75,ymin=0,ymax=2] {Figures/T2_nSVD_20.png};
	\end{axis}
	
	\begin{axis}[%
	width=4cm,
	height=4cm,
	axis on top,
	scale only axis,
	xmin=0,
	xmax=1,
	y dir=reverse,
	ymin=0,
	ymax=1,
	hide axis,
	name=T2ADMM,
	colormap name = morgenstemning,
	colorbar,
	point meta min=0,
	point meta max=0.2,
	at=(T2SVD.right of north east),
	anchor=left of north west,
	xshift=.08in,
	colorbar style={ylabel=$T_2~(\text{s})$, width=0.5cm, xshift=-0.1cm, yticklabel style={/pgf/number format/fixed}},
	]
	\addplot graphics [xmin=-0.25,xmax=1.75,ymin=0,ymax=2] {Figures/T2_ADMM_nSVD_20.png};
	\end{axis}
	\end{tikzpicture}
	\fi
	\caption{\CaptMapsT}
	\label{fig:Maps_nSVD_20}
\end{figure}

\begin{figure}[hp]
	\centering
	\if\submit1
	\includegraphics{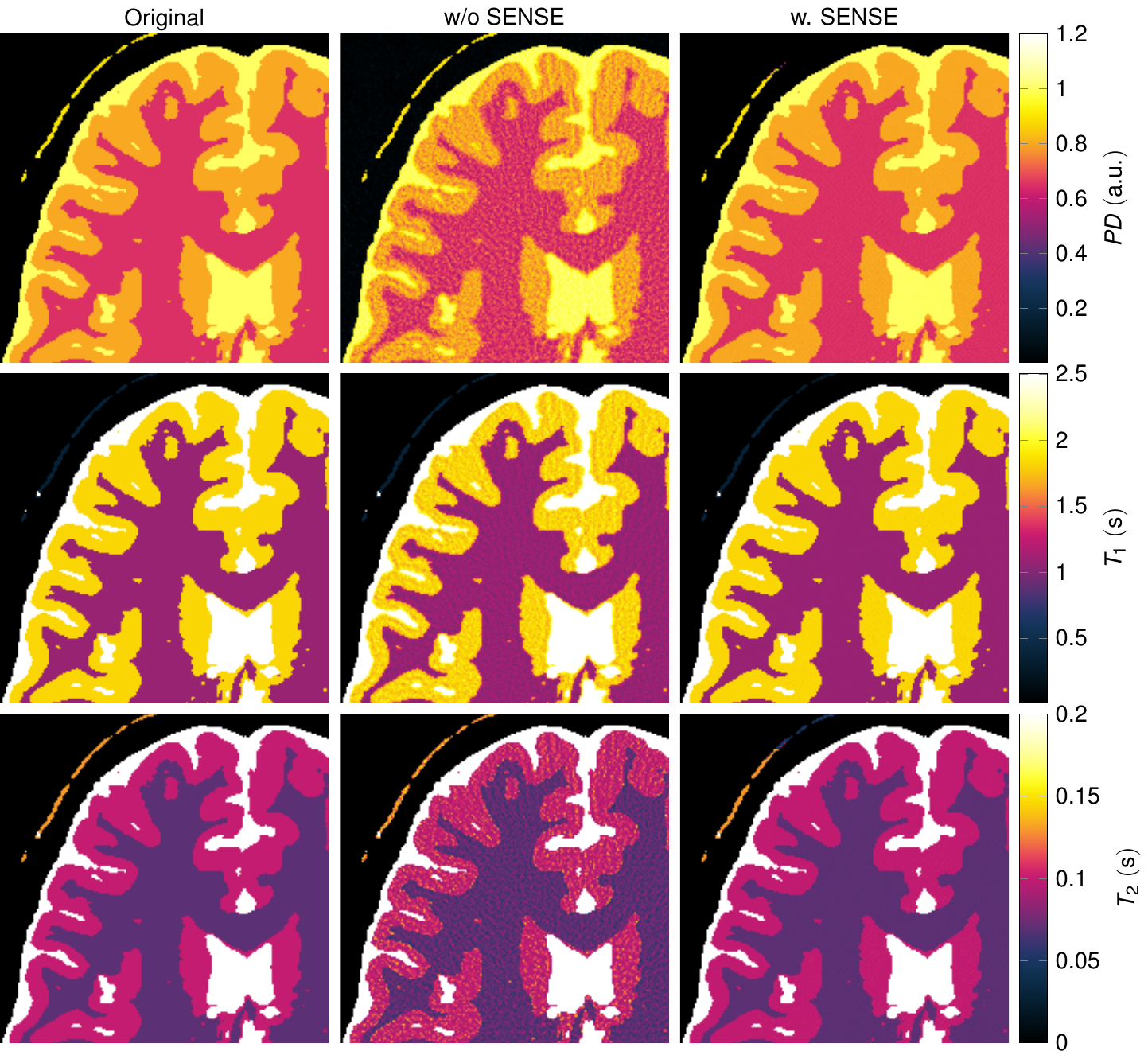}
	\else
	
	\begin{tikzpicture}[scale=.7]
	\begin{axis}[%
	width=6cm,
	height=6cm,
	axis on top,
	scale only axis,
	xmin=0,
	xmax=1,
	y dir=reverse,
	ymin=0,
	ymax=1,
	hide axis,
	name=PDOrg,
	title={Original},
	title style={yshift = -.3cm},
	]
	\addplot graphics [xmin=-0.25,xmax=1.75,ymin=0,ymax=2] {Figures/PD_SENSE_org.png};
	\end{axis}
	
	\begin{axis}[%
	width=6cm,
	height=6cm,
	axis on top,
	scale only axis,
	xmin=0,
	xmax=1,
	y dir=reverse,
	ymin=0,
	ymax=1,
	hide axis,
	name=PDMRF,
	title={w/o SENSE},
	title style={yshift = -.2cm},
	at=(PDOrg.right of north east),
	anchor=left of north west,
	xshift=.08in,
	]
	\addplot graphics [xmin=-0.25,xmax=1.75,ymin=0,ymax=2] {Figures/PD_ADMM_HighRes.png};
	\end{axis}
	
	\begin{axis}[%
	width=6cm,
	height=6cm,
	axis on top,
	scale only axis,
	xmin=0,
	xmax=1,
	y dir=reverse,
	ymin=0,
	ymax=1,
	hide axis,
	name=PDSVDBP,
	title={w. SENSE},
	title style={yshift = -.2cm},
	colormap name = morgenstemning,
	colorbar,
	point meta min=0.001,
	point meta max=1.2,
	at=(PDMRF.right of north east),
	anchor=left of north west,
	xshift=.08in,
	colorbar style={ylabel=$PD~(\text{a.u.})$, width=0.5cm, xshift=-0.1cm},
	]
	\addplot graphics [xmin=-0.25,xmax=1.75,ymin=0,ymax=2] {Figures/PD_ADMM_SENSE.png};
	\end{axis}
	
	\begin{axis}[%
	width=6cm,
	height=6cm,
	axis on top,
	scale only axis,
	xmin=0,
	xmax=1,
	y dir=reverse,
	ymin=0,
	ymax=1,
	hide axis,
	name=T1Org,
	colormap name = morgenstemning,
	at=(PDOrg.south east),
	anchor=north east,
	yshift = -.08in,
	]
	\addplot graphics [xmin=-0.25,xmax=1.75,ymin=0,ymax=2] {Figures/T1_SENSE_org.png};
	\end{axis}
	
	\begin{axis}[%
	width=6cm,
	height=6cm,
	axis on top,
	scale only axis,
	xmin=0,
	xmax=1,
	y dir=reverse,
	ymin=0,
	ymax=1,
	hide axis,
	name=T1MRF,
	colormap name = morgenstemning,
	point meta min=0.01,
	point meta max=1.5,
	at=(T1Org.right of north east),
	anchor=left of north west,
	xshift=.08in,
	]
	\addplot graphics [xmin=-0.25,xmax=1.75,ymin=0,ymax=2] {Figures/T1_ADMM_HighRes.png};
	\end{axis}
	
	\begin{axis}[%
	width=6cm,
	height=6cm,
	axis on top,
	scale only axis,
	xmin=0,
	xmax=1,
	y dir=reverse,
	ymin=0,
	ymax=1,
	hide axis,
	name=T1SVDBP,
	colormap name = morgenstemning,
	colorbar,
	point meta min=0.01,
	point meta max=2.5,
	at=(T1MRF.right of north east),
	anchor=left of north west,
	xshift=.08in,
	colorbar style={ylabel=$T_1~(\text{s})$, width=0.5cm, xshift=-0.1cm},
	]
	\addplot graphics [xmin=-0.25,xmax=1.75,ymin=0,ymax=2] {Figures/T1_ADMM_SENSE.png};
	\end{axis}
	
	\begin{axis}[%
	width=6cm,
	height=6cm,
	axis on top,
	scale only axis,
	xmin=0,
	xmax=1,
	y dir=reverse,
	ymin=0,
	ymax=1,
	hide axis,
	name=T2Org,
	colormap name = morgenstemning,
	at=(T1Org.south east),
	anchor=north east,
	yshift = -.08in,
	]
	\addplot graphics [xmin=-0.25,xmax=1.75,ymin=0,ymax=2] {Figures/T2_SENSE_org.png};
	\end{axis}
	
	\begin{axis}[%
	width=6cm,
	height=6cm,
	axis on top,
	scale only axis,
	xmin=0,
	xmax=1,
	y dir=reverse,
	ymin=0,
	ymax=1,
	hide axis,
	name=T2MRF,
	colormap name = morgenstemning,
	point meta min=0,
	point meta max=0.3,
	at=(T2Org.right of north east),
	anchor=left of north west,
	xshift=.08in,
	]
	\addplot graphics [xmin=-0.25,xmax=1.75,ymin=0,ymax=2] {Figures/T2_ADMM_HighRes.png};
	\end{axis}
	
	\begin{axis}[%
	width=6cm,
	height=6cm,
	axis on top,
	scale only axis,
	xmin=0,
	xmax=1,
	y dir=reverse,
	ymin=0,
	ymax=1,
	hide axis,
	name=T2SVDBP,
	colormap name = morgenstemning,
	colorbar,
	point meta min=0,
	point meta max=0.2,
	at=(T2MRF.right of north east),
	anchor=left of north west,
	xshift=.08in,
	colorbar style={ylabel=$T_2~(\text{s})$, width=0.5cm, xshift=-0.1cm, yticklabel style={/pgf/number format/fixed},},
	]
	\addplot graphics [xmin=-0.25,xmax=1.75,ymin=0,ymax=2] {Figures/T2_ADMM_SENSE.png};
	\end{axis}
	\end{tikzpicture}
	\fi
	\caption{\CaptSENSE}
	\label{fig:SENSE}
\end{figure}

\begin{figure}[hp]
	\centering
	\if\submit1
	\includegraphics{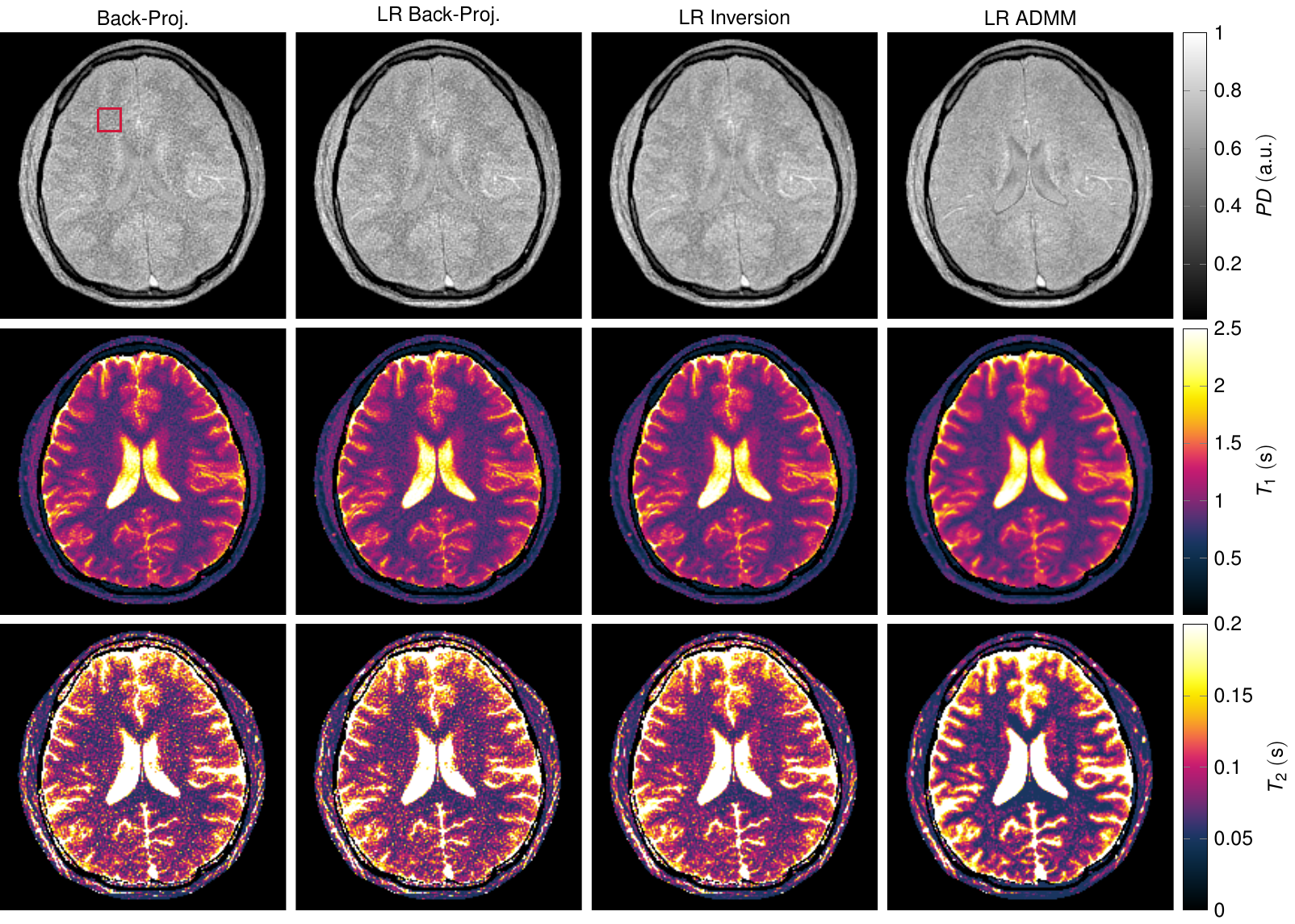}
	\else
	
	\begin{tikzpicture}[scale=.6]
	\begin{axis}[%
	width=6cm,
	height=6cm,
	axis on top,
	scale only axis,
	xmin=0,
	xmax=1,
	y dir=reverse,
	ymin=0,
	ymax=1,
	hide axis,
	name=PDBP,
	colormap/blackwhite,
	point meta min=0.01,
	point meta max=1,	
	colorbar style={ylabel=$PD~[\text{a.u.}]$, width=0.5cm, xshift=-0.1cm},
	title={Back-Proj.},
	title style={yshift = -.3cm},
	]
	\addplot graphics [xmin=0,xmax=1,ymin=0,ymax=1] {Figures/InVivo_PD_BW_BP_1_spoke.png};
	
	\addplot[UKLred, no marks, ultra  thick] coordinates {(66/192,51/192) (66/192,66/192) (81/192,66/192) (81/192,51/192) (66/192,51/192)};
	\end{axis}
	
	\begin{axis}[%
	width=6cm,
	height=6cm,
	axis on top,
	scale only axis,
	xmin=0,
	xmax=1,
	y dir=reverse,
	ymin=0,
	ymax=1,
	hide axis,
	name=T1BP,
	colormap name = morgenstemning,
	point meta min=0.01,
	point meta max=2.5,
	at=(PDBP.below south west),
	anchor=above north west,
	yshift=-.08in,
	colorbar style={ylabel=$T_1~[\text{s}]$, width=0.5cm, xshift=-0.1cm},
	]
	\addplot graphics [xmin=0,xmax=1,ymin=0,ymax=1] {Figures/InVivo_T1_BP_1_spoke.png};
	\end{axis}
	
	\begin{axis}[%
	width=6cm,
	height=6cm,
	axis on top,
	scale only axis,
	xmin=0,
	xmax=1,
	y dir=reverse,
	ymin=0,
	ymax=1,
	hide axis,
	name=T2BP,
	colormap name = morgenstemning,
	point meta min=0,
	point meta max=.25,
	at=(T1BP.below south west),
	anchor=above north west,
	yshift=-.08in,
	colorbar style={ylabel=$T_2~[\text{s}]$, width=0.5cm, xshift=-0.1cm, yticklabel style={/pgf/number format/fixed},},
	]
	\addplot graphics [xmin=0,xmax=1,ymin=0,ymax=1] {Figures/InVivo_T2_BP_1_spoke.png};
	\end{axis}
	
	\begin{axis}[%
	width=6cm,
	height=6cm,
	axis on top,
	scale only axis,
	xmin=0,
	xmax=1,
	y dir=reverse,
	ymin=0,
	ymax=1,
	hide axis,
	name=PDLRBP,
	at=(PDBP.right of north east),
	anchor= left of north west,
	xshift=.08in,
	colormap/blackwhite,
	point meta min=0.01,
	point meta max=1,	
	title={LR Back-Proj.},
	title style={yshift = -.2cm},
	]
	\addplot graphics [xmin=0,xmax=1,ymin=0,ymax=1] {Figures/InVivo_PD_BW_LR_BP_1_spoke.png};
	\end{axis}
	
	\begin{axis}[%
	width=6cm,
	height=6cm,
	axis on top,
	scale only axis,
	xmin=0,
	xmax=1,
	y dir=reverse,
	ymin=0,
	ymax=1,
	hide axis,
	name=T1LRBP,
	colormap name = morgenstemning,
	point meta min=0.01,
	point meta max=2.5,
	at=(PDLRBP.below south west),
	anchor=above north west,
	yshift=-.08in,
	]
	\addplot graphics [xmin=0,xmax=1,ymin=0,ymax=1] {Figures/InVivo_T1_LR_BP_1_spoke.png};
	\end{axis}
	
	\begin{axis}[%
	width=6cm,
	height=6cm,
	axis on top,
	scale only axis,
	xmin=0,
	xmax=1,
	y dir=reverse,
	ymin=0,
	ymax=1,
	hide axis,
	name=T2LRBP,
	colormap name = morgenstemning,
	point meta min=0,
	point meta max=.2,
	at=(T1LRBP.below south west),
	anchor=above north west,
	yshift=-.08in,
	]
	\addplot graphics [xmin=0,xmax=1,ymin=0,ymax=1] {Figures/InVivo_T2_LR_BP_1_spoke.png};
	\end{axis}
	
	\begin{axis}[%
	width=6cm,
	height=6cm,
	axis on top,
	scale only axis,
	xmin=0,
	xmax=1,
	y dir=reverse,
	ymin=0,
	ymax=1,
	hide axis,
	name=PDLRInv,
	at=(PDLRBP.right of north east),
	anchor= left of north west,
	xshift=.08in,
	colormap/blackwhite,
	point meta min=0.01,
	point meta max=1,	
	title={LR Inversion},
	title style={yshift = -.2cm},
	]
	\addplot graphics [xmin=0,xmax=1,ymin=0,ymax=1] {Figures/InVivo_PD_Inv_BW_1_spoke.png};
	\end{axis}
	
	\begin{axis}[%
	width=6cm,
	height=6cm,
	axis on top,
	scale only axis,
	xmin=0,
	xmax=1,
	y dir=reverse,
	ymin=0,
	ymax=1,
	hide axis,
	name=T1LRInv,
	colormap name = morgenstemning,
	point meta min=0.01,
	point meta max=2.5,
	at=(PDLRInv.below south west),
	anchor=above north west,
	yshift=-.08in,
	]
	\addplot graphics [xmin=0,xmax=1,ymin=0,ymax=1] {Figures/InVivo_T1_Inv_1_spoke.png};
	\end{axis}
	
	\begin{axis}[%
	width=6cm,
	height=6cm,
	axis on top,
	scale only axis,
	xmin=0,
	xmax=1,
	y dir=reverse,
	ymin=0,
	ymax=1,
	hide axis,
	name=T2LRInv,
	colormap name = morgenstemning,
	point meta min=0,
	point meta max=.2,
	at=(T1LRInv.below south west),
	anchor=above north west,
	yshift=-.08in,
	]
	\addplot graphics [xmin=0,xmax=1,ymin=0,ymax=1] {Figures/InVivo_T2_Inv_1_spoke.png};
	\end{axis}

	\begin{axis}[%
	width=6cm,
	height=6cm,
	axis on top,
	scale only axis,
	xmin=0,
	xmax=1,
	y dir=reverse,
	ymin=0,
	ymax=1,
	hide axis,
	name=PDADMM,
	at=(PDLRInv.right of north east),
	anchor= left of north west,
	xshift=.08in,
	colormap/blackwhite,
	colorbar,
	point meta min=0.01,
	point meta max=1,	
	colorbar style={ylabel=$PD~(\text{a.u.})$, width=0.5cm, xshift=-0.1cm},
	title={LR ADMM},
	title style={yshift = -.2cm},
	]
	\addplot graphics [xmin=0,xmax=1,ymin=0,ymax=1] {Figures/InVivo_PD_BW_1_spoke.png};
	\end{axis}
	
	\begin{axis}[%
	width=6cm,
	height=6cm,
	axis on top,
	scale only axis,
	xmin=0,
	xmax=1,
	y dir=reverse,
	ymin=0,
	ymax=1,
	hide axis,
	name=T1ADMM,
	colormap name = morgenstemning,
	colorbar,
	point meta min=0.01,
	point meta max=2.5,
	at=(PDADMM.below south west),
	anchor=above north west,
	yshift=-.08in,
	colorbar style={ylabel=$T_1~(\text{s})$, width=0.5cm, xshift=-0.1cm},
	]
	\addplot graphics [xmin=0,xmax=1,ymin=0,ymax=1] {Figures/InVivo_T1_1_spoke.png};
	\end{axis}
	
	\begin{axis}[%
	width=6cm,
	height=6cm,
	axis on top,
	scale only axis,
	xmin=0,
	xmax=1,
	y dir=reverse,
	ymin=0,
	ymax=1,
	hide axis,
	name=T2ADMM,
	colormap name = morgenstemning,
	colorbar,
	point meta min=0,
	point meta max=.2,
	at=(T1ADMM.below south west),
	anchor=above north west,
	yshift=-.08in,
	colorbar style={ylabel=$T_2~(\text{s})$, width=0.5cm, xshift=-0.1cm, yticklabel style={/pgf/number format/fixed},},
	]
	\addplot graphics [xmin=0,xmax=1,ymin=0,ymax=1] {Figures/InVivo_T2_1_spoke.png};
	\end{axis}
	\end{tikzpicture}
	\fi
	\caption{\CaptInVivo}
	\label{fig:InVivo}
\end{figure}

\begin{figure}[hp]
\centering
\if\submit1
\includegraphics{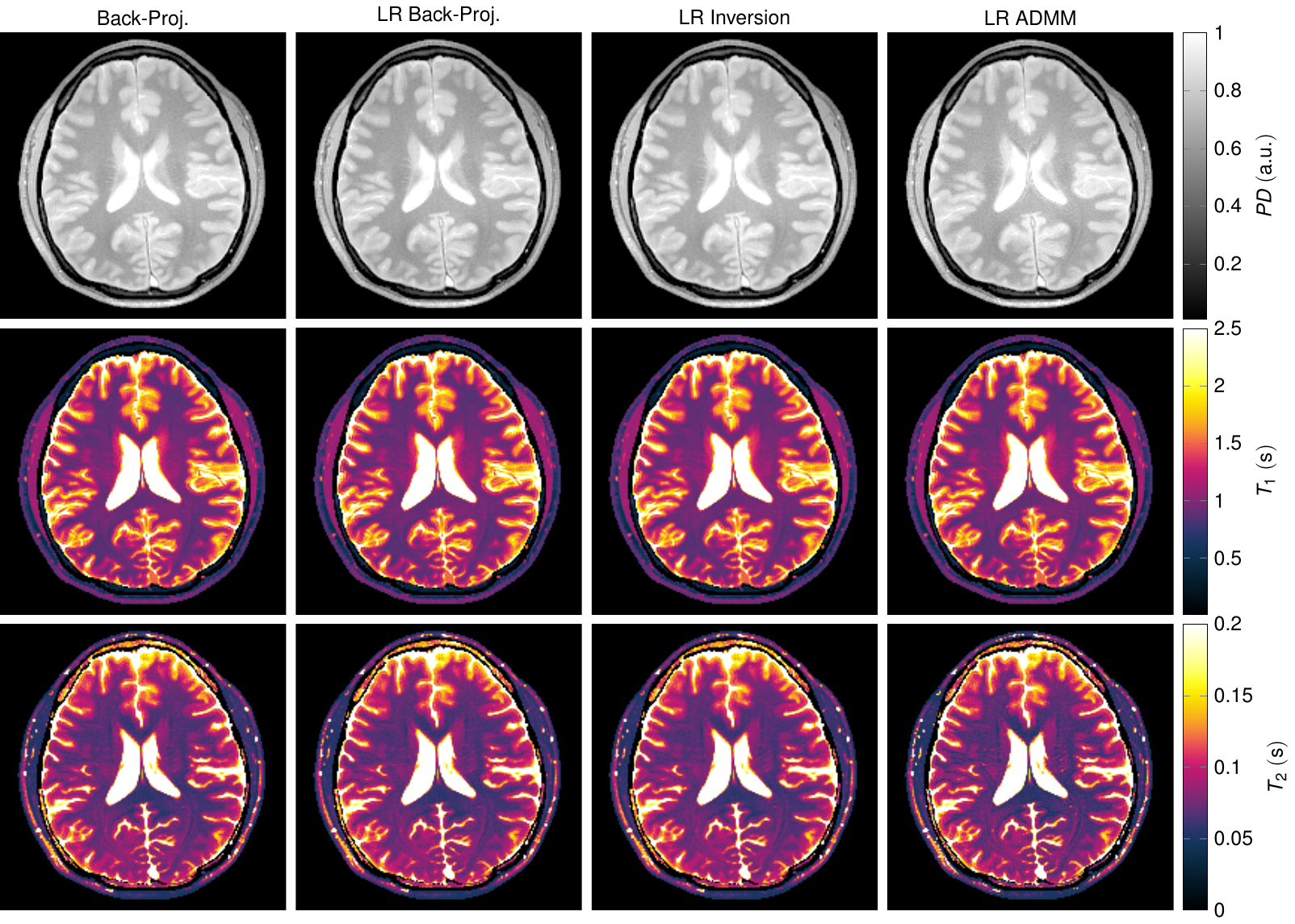}
\else
\begin{tikzpicture}[scale=.6]
\begin{axis}[%
width=6cm,
height=6cm,
axis on top,
scale only axis,
xmin=0,
xmax=1,
y dir=reverse,
ymin=0,
ymax=1,
hide axis,
name=PDBP,
colormap/blackwhite,
point meta min=0.01,
point meta max=1,	
colorbar style={ylabel=$PD~[\text{a.u.}]$, width=0.5cm, xshift=-0.1cm},
title={Back-Proj.},
title style={yshift = -.3cm},
]
\addplot graphics [xmin=0,xmax=1,ymin=0,ymax=1] {Figures/InVivo_PD_BW_BP_32_spoke.png};
\end{axis}

\begin{axis}[%
width=6cm,
height=6cm,
axis on top,
scale only axis,
xmin=0,
xmax=1,
y dir=reverse,
ymin=0,
ymax=1,
hide axis,
name=T1BP,
colormap name = morgenstemning,
point meta min=0.01,
point meta max=2.5,
at=(PDBP.below south west),
anchor=above north west,
yshift=-.08in,
colorbar style={ylabel=$T_1~[\text{s}]$, width=0.5cm, xshift=-0.1cm},
]
\addplot graphics [xmin=0,xmax=1,ymin=0,ymax=1] {Figures/InVivo_T1_BP_32_spoke.png};
\end{axis}

\begin{axis}[%
width=6cm,
height=6cm,
axis on top,
scale only axis,
xmin=0,
xmax=1,
y dir=reverse,
ymin=0,
ymax=1,
hide axis,
name=T2BP,
colormap name = morgenstemning,
point meta min=0,
point meta max=.25,
at=(T1BP.below south west),
anchor=above north west,
yshift=-.08in,
colorbar style={ylabel=$T_2~[\text{s}]$, width=0.5cm, xshift=-0.1cm, yticklabel style={/pgf/number format/fixed},},
]
\addplot graphics [xmin=0,xmax=1,ymin=0,ymax=1] {Figures/InVivo_T2_BP_32_spoke.png};
\end{axis}

\begin{axis}[%
width=6cm,
height=6cm,
axis on top,
scale only axis,
xmin=0,
xmax=1,
y dir=reverse,
ymin=0,
ymax=1,
hide axis,
name=PDLRBP,
at=(PDBP.right of north east),
anchor= left of north west,
xshift=.08in,
colormap/blackwhite,
point meta min=0.01,
point meta max=1,	
title={LR Back-Proj.},
title style={yshift = -.2cm},
]
\addplot graphics [xmin=0,xmax=1,ymin=0,ymax=1] {Figures/InVivo_PD_BW_LR_BP_32_spoke.png};
\end{axis}

\begin{axis}[%
width=6cm,
height=6cm,
axis on top,
scale only axis,
xmin=0,
xmax=1,
y dir=reverse,
ymin=0,
ymax=1,
hide axis,
name=T1LRBP,
colormap name = morgenstemning,
point meta min=0.01,
point meta max=2.5,
at=(PDLRBP.below south west),
anchor=above north west,
yshift=-.08in,
]
\addplot graphics [xmin=0,xmax=1,ymin=0,ymax=1] {Figures/InVivo_T1_LR_BP_32_spoke.png};
\end{axis}

\begin{axis}[%
width=6cm,
height=6cm,
axis on top,
scale only axis,
xmin=0,
xmax=1,
y dir=reverse,
ymin=0,
ymax=1,
hide axis,
name=T2LRBP,
colormap name = morgenstemning,
point meta min=0,
point meta max=.2,
at=(T1LRBP.below south west),
anchor=above north west,
yshift=-.08in,
]
\addplot graphics [xmin=0,xmax=1,ymin=0,ymax=1] {Figures/InVivo_T2_LR_BP_32_spoke.png};
\end{axis}

\begin{axis}[%
width=6cm,
height=6cm,
axis on top,
scale only axis,
xmin=0,
xmax=1,
y dir=reverse,
ymin=0,
ymax=1,
hide axis,
name=PDLRInv,
at=(PDLRBP.right of north east),
anchor= left of north west,
xshift=.08in,
colormap/blackwhite,
point meta min=0.01,
point meta max=1,	
title={LR Inversion},
title style={yshift = -.2cm},
]
\addplot graphics [xmin=0,xmax=1,ymin=0,ymax=1] {Figures/InVivo_PD_Inv_BW_32_spoke.png};
\end{axis}

\begin{axis}[%
width=6cm,
height=6cm,
axis on top,
scale only axis,
xmin=0,
xmax=1,
y dir=reverse,
ymin=0,
ymax=1,
hide axis,
name=T1LRInv,
colormap name = morgenstemning,
point meta min=0.01,
point meta max=2.5,
at=(PDLRInv.below south west),
anchor=above north west,
yshift=-.08in,
]
\addplot graphics [xmin=0,xmax=1,ymin=0,ymax=1] {Figures/InVivo_T1_Inv_32_spoke.png};
\end{axis}

\begin{axis}[%
width=6cm,
height=6cm,
axis on top,
scale only axis,
xmin=0,
xmax=1,
y dir=reverse,
ymin=0,
ymax=1,
hide axis,
name=T2LRInv,
colormap name = morgenstemning,
point meta min=0,
point meta max=.2,
at=(T1LRInv.below south west),
anchor=above north west,
yshift=-.08in,
]
\addplot graphics [xmin=0,xmax=1,ymin=0,ymax=1] {Figures/InVivo_T2_Inv_32_spoke.png};
\end{axis}

\begin{axis}[%
width=6cm,
height=6cm,
axis on top,
scale only axis,
xmin=0,
xmax=1,
y dir=reverse,
ymin=0,
ymax=1,
hide axis,
name=PDADMM,
at=(PDLRInv.right of north east),
anchor= left of north west,
xshift=.08in,
colormap/blackwhite,
colorbar,
point meta min=0.01,
point meta max=1,	
colorbar style={ylabel=$PD~(\text{a.u.})$, width=0.5cm, xshift=-0.1cm},
title={LR ADMM},
title style={yshift = -.2cm},
]
\addplot graphics [xmin=0,xmax=1,ymin=0,ymax=1] {Figures/InVivo_PD_BW_32_spoke.png};
\end{axis}

\begin{axis}[%
width=6cm,
height=6cm,
axis on top,
scale only axis,
xmin=0,
xmax=1,
y dir=reverse,
ymin=0,
ymax=1,
hide axis,
name=T1ADMM,
colormap name = morgenstemning,
colorbar,
point meta min=0.01,
point meta max=2.5,
at=(PDADMM.below south west),
anchor=above north west,
yshift=-.08in,
colorbar style={ylabel=$T_1~(\text{s})$, width=0.5cm, xshift=-0.1cm},
]
\addplot graphics [xmin=0,xmax=1,ymin=0,ymax=1] {Figures/InVivo_T1_32_spoke.png};
\end{axis}

\begin{axis}[%
width=6cm,
height=6cm,
axis on top,
scale only axis,
xmin=0,
xmax=1,
y dir=reverse,
ymin=0,
ymax=1,
hide axis,
name=T2ADMM,
colormap name = morgenstemning,
colorbar,
point meta min=0,
point meta max=.2,
at=(T1ADMM.below south west),
anchor=above north west,
yshift=-.08in,
colorbar style={ylabel=$T_2~(\text{s})$, width=0.5cm, xshift=-0.1cm, yticklabel style={/pgf/number format/fixed},},
]
\addplot graphics [xmin=0,xmax=1,ymin=0,ymax=1] {Figures/InVivo_T2_32_spoke.png};
\end{axis}
\end{tikzpicture}
\fi
\caption{\CaptInVivoSpokes}
\label{fig:InVivo_32spokes}
\end{figure}

\clearpage 
\begin{table}[h!]
	\centering
	\begin{tabular}{|c|c|c|c|c|}
		\hline
		Repetitions        	& 1 & 32 & 1 & 32 \\ \hline
				                    & \multicolumn{2}{|c|}{$T_1~\text{(ms)}$} & \multicolumn{2}{|c|}{$T_2~\text{(ms)}$} \\ \hline
		Back-Proj.			& $835\pm99$  & $909\pm48$ & $72\pm20$ & $70\pm5$ \\ \hline
		LR Back-Proj.	 & $835\pm100$ & $909\pm48$ & $73\pm20$ & $70\pm5$  \\ \hline
		LR Inversion      & $838\pm81$   & $917\pm45$ & $71\pm13$ & $71\pm4$  \\ \hline
		LR ADMM			 & $826\pm62$  & $917\pm48$ & $64\pm11$ & $70\pm6$  \\ \hline
		Literature \cite{Jiang2014a}& \multicolumn{2}{|c|}{$781\pm61$}   & \multicolumn{2}{|c|}{$65\pm6$}  \\ \hline
	\end{tabular}
	\caption{\CaptInVivoTable}
	\label{tab:in_vivo}
\end{table}
\end{document}